\documentclass[twocolumn,prb,showpacs,preprintnumbers,superscriptaddress,amsmath,amssymb,citeautoscript]{revtex4-1}

\usepackage{graphicx}
\usepackage{epstopdf}
\usepackage{dcolumn}
\usepackage{color}
\usepackage{multirow}
\usepackage{rotating}
\usepackage{array}
\usepackage{bm}

\DeclareMathOperator{\sgn}{sgn}

\newcommand{\dd}{{\mathbf d}}
\newcommand{\be}{\begin{equation}}
\newcommand{\ee}{\end{equation}}
\newcommand{\ba}{\begin{eqnarray}}
\newcommand{\ea}{\end{eqnarray}}
\newcommand{\bpm}{\begin{pmatrix}}
\newcommand{\epm}{\end{pmatrix}}

\begin{document}

\title{Characterizing unconventional superconductors from the spin structure of impurity-induced bound states}

\author{V. Kaladzhyan}
\affiliation{Institut de Physique Th\'eorique, CEA/Saclay,
Orme des Merisiers, 91190 Gif-sur-Yvette Cedex, France}
\affiliation{Laboratoire de Physique des Solides, CNRS, Univ. Paris-Sud, Universit\'e Paris-Saclay, 91405 Orsay Cedex, France}
\author{C. Bena}
\affiliation{Institut de Physique Th\'eorique, CEA/Saclay,
Orme des Merisiers, 91190 Gif-sur-Yvette Cedex, France}
\affiliation{Laboratoire de Physique des Solides, CNRS, Univ. Paris-Sud, Universit\'e Paris-Saclay, 91405 Orsay Cedex, France}
\author{P. Simon}
\affiliation{Laboratoire de Physique des Solides, CNRS, Univ. Paris-Sud, Universit\'e Paris-Saclay, 91405 Orsay Cedex, France}

\date{\today}

\begin{abstract}
 Cooper pairs in two-dimensional unconventional superconductors with broken inversion symmetry 
are in a mixture of an even-parity spin-singlet pairing state   with an odd-parity spin-triplet pairing state.
We study the magnetic properties of the impurity bound states in such superconductors and find striking
signatures in  their spin polarization which allow to  unambiguously discriminate a non-topological superconducting phase from a topological one. Moreover, we show how these properties, which could be measured using spin-polarized scanning 
tunneling microscopy (STM),  also enable to determine the direction of the spin-triplet pairing vector of the host material and thus to distinguish between different types of unconventional pairing.
\end{abstract}

\maketitle

\section{Introduction}
Ever since the discovery of strontium ruthenate as one the first unconventional
superconductors (SCs) more than thirty years ago \cite{Maeno2003}, the search for the symmetry of Cooper pairs \cite{Mineev1999,Sigrist2009}
has been among the most important tasks to be addressed in order to characterize new
superconducting  materials. 
In the past decade, many new SCs with broken inversion symmetry have been discovered.
These SCs are expected to display unconventional
pairing due to the strong  spin-orbit coupling. Indeed, in these materials, 
the twofold spin degeneracy is lifted by spin-orbit interaction, and the Cooper pairs exhibit a mixture of singlet and triplet pairing \cite{Gorkov-Rashba,sigrist2004}.
Examples of such systems can be found in non-centrosymmetric SCs \cite{Sigrist2009,Bauer2012} and doped topological insulators \cite{Sasaki2011}. Other materials with unconventional pairing can also be uncovered within  two-dimensional (2D) or quasi-2D superconductors which necessarily break the 3D inversion symmetry, such as the surface states of topological insulators in the proximity of s-wave superconductors \cite{Fu-Kane},  2D materials with strong  spin-orbit coupling proximitized by s-wave superconductivity \cite{Sau2010,Alicea2010}, 
monolayer or  few-layer transition metal dichalcogenides \cite{Lu2015,Xi2015}, etc. 

The fact that the surface states of non-centrosymmetric SCs  with mixed singlet and triplet pairing are predicted to support spin-polarized currents \cite{Eschrig2008,Tanaka2009} suggests that such 2D unconventional superconductors may exhibit a non-trivial spin response to local magnetic fields and magnetic impurities. 
In s-wave SCs, magnetic impurities lead to  so-called intra-gap Shiba bound states (SBSs) \cite{Yu1965,Shiba1968,Rusinov1969} (see \cite{Balatsky2006} for a review) which have experimentally been probed using scanning tunneling microscopy (STM) \cite{Yazdani1997, Shuai-Hua2008,Franke2015,Menard2015}.


While a point-like scalar impurity does not induce a SBS in s-wave SCs \cite{Balatsky2006}, 
it gives rise to one SBS in p-wave dominant SCs \cite{Wang2004,Eremin2008,Nagai2014,Lutchyn2015}.
More interestingly, a point-like magnetic impurity leads to
the formation of one SBS in the s-dominant regime and of two SBSs in the p-dominant regime, although as we show here, not all of them are always subgap states. 
This may suggest to use the number of SBSs as a natural criterion to discriminate between s-wave and p-wave SCs. However, we note that this quantity  depends  on the number  of impurity orbitals which hybridyze with the SC as well as the impurity shape.  Recently it has also been shown that  Shiba states in p-wave SCs have a non-trivial spectral dependence with respect to the spin-orbit coupling (SOC) or with the direction of the magnetic moment  \cite{Lutchyn2015}, compared to the s-wave SCs. 
However, the former parameter (SOC) is usually given while the latter is not  so easy to handle experimentally with magnetic fields.

In this paper, we propose to determine the degree of triplet pairing together with the orientation 
with respect to the sample plane 
of the spin-triplet pairing vector (the so-called $\dd$ vector \cite{Mineev1999,Sigrist2009})  using spin-polarized STM measurements of the integrated spin-polarized local density of states (SP LDOS) of the SBSs,
 as well as its Fourier transform (FT), for the Shiba states associated with magnetic impurities. 
We find that for an impurity with a spin orthogonal to the $\dd$ vector, 
the particle and hole component of the SBSs closest to midgap have spins of the same sign in the p-dominant regime, and of opposite sign in the s-dominant regime. Furthermore, when both the $\dd$ vector and the impurity spin are in-plane, we find a spectacular cancellation of half of the  in-plane spin components of the SBSs which can be traced  back 
to the orbital pairing nature of the host SC. 
Therefore, the relative sign of the spin for these two states can serve as a probe to test experimentally the degree 
of p-wave pairing via spin-polarized STM measurements or using spin-polarized transport experiments.
%
%
%
Moreover, by studying the Fourier transform (FT) of the electronic SP LDOS of these states, 
we find  distinct features, the most striking being that the FT of the SP LDOS acquires a four-fold-symmetry, characteristic for the orbital p-wave pairing, in the topological p-wave dominant regime when both the $\dd$ vector and the impurity spin are in-plane. 


\begin{figure}[h]
\noindent\includegraphics[width=\columnwidth]{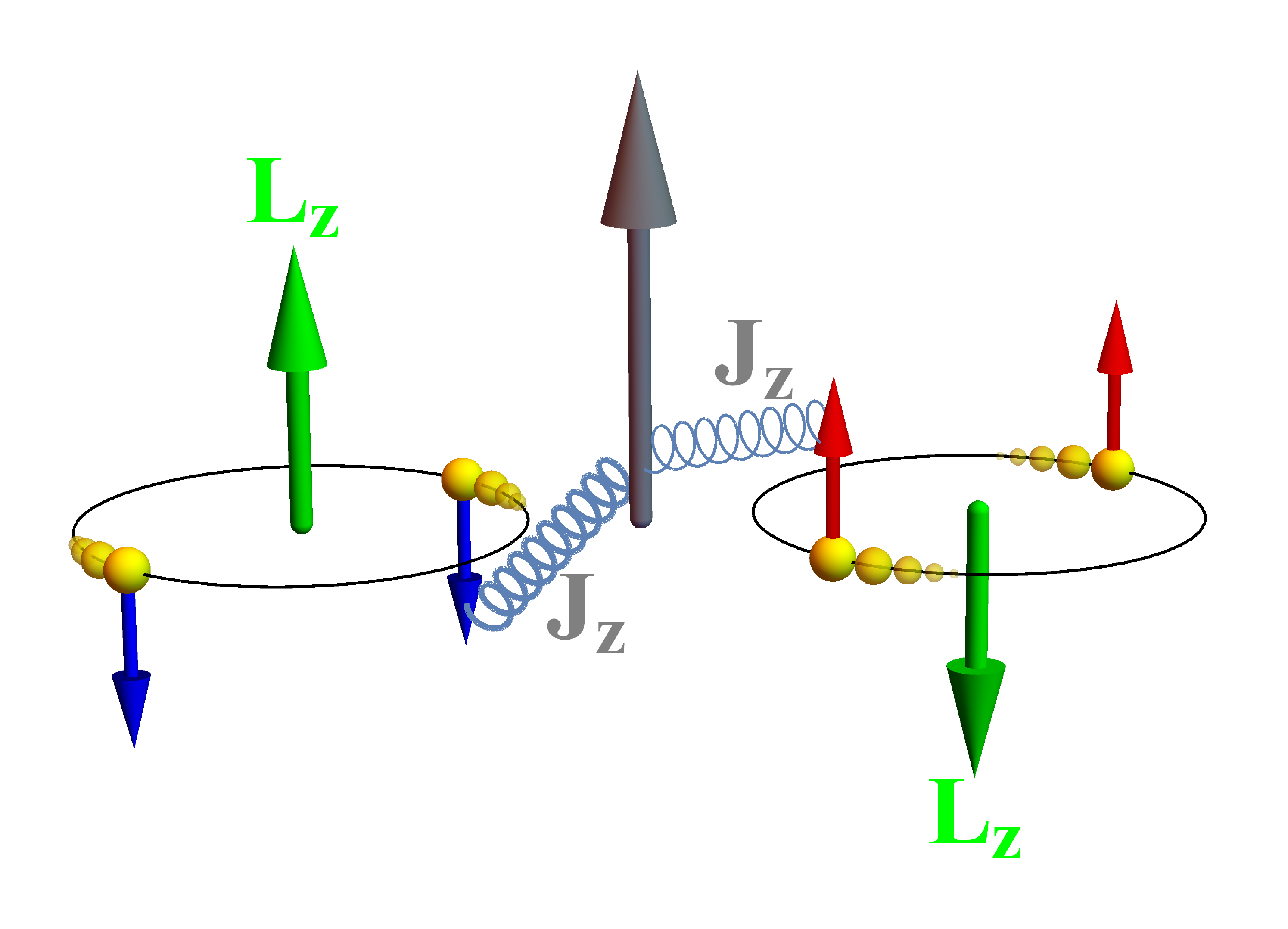}
	\caption{(Color online) Sketch of an out-of-plane magnetic impurity in a p-wave TRS superconductor with an in-plane $\dd$ vector. In this case, electrons forming Cooper pairs have always some out-of-plane spin components  (blue and red arrows) \cite{Maeno2003}. Furthermore, due to TRS, Cooper pairs  have either an angular momentum (green arrow) with $L_z=+1$ and are composed of electrons with spins $s_z=-1/2$ or the opposite \cite{Maeno2003}. TRS is locally broken by a magnetic impurity. For antiferromagnetic exchange interactions, $J_z>0$, breaking Cooper pairs with  electron spins $s_z=-1/2$ (blue arrows) is energetically favoured compared to Cooper pairs with spins $s_z=1/2$ (red arrows), hence two non-degenerate in-gap bound states are expected.
}
	\label{Sketch}
\end{figure}

\section{Model}
We consider a 2D SC with both s-wave SC pairing $\Delta_s$ and p-wave SC pairing $\varkappa$. The corresponding Hamiltonian can be written in the Nambu basis $\Psi_{\bm{k}}=(\psi_{\uparrow \bm{k}},\psi_{\downarrow \bm{k}},\psi^{\dag}_{\downarrow -\bm{k}},-\psi^{\dag}_{\uparrow -\bm{k}})^\mathrm{T}$ as
\ba
\mathcal{H}_0 =	
	\begin{pmatrix}
		\Xi_{\bm{k}} \sigma_0 & \Delta(\bm{k}) \\
		\Delta^\dag(\bm{k}) & -\Xi_{\bm{k}} \sigma_0 \\
	\end{pmatrix}.	
\label{h0}
\ea	
where $\Delta(\bm{k}) = \Delta_s \sigma_0 + \varkappa\, \bm{d}(\bm{k}) \cdot \bm{\sigma}$ is the pairing function. 
In order to simplify the discussion, we assume in what follows that
$\Delta_s$ and $\varkappa$ are real and positive.
 The symbol ${\bm\sigma}$ denotes the Pauli matrices acting in the spin subspace. 
The operator $\psi^\dag_{\sigma {\bm k}}$ creates a particle of spin $\sigma=\uparrow,\downarrow$ of momentum ${\bm k}=(k_x,k_y)$. We  set  the Fermi velocity $v_F=1$ and $\hbar=1$. The system is considered to lay in the $(x,y)$ plane. The energy dispersion in the normal state is embodied by $\Xi_{\bm k} $. Since our main message is barely modified by the presence of the spin-orbit coupling term, we do not consider it in the main text but discuss its effect in the supplementary information (SI) \cite{sm}, whereas its effect on the spin polarization of SBSs in superconductors with purely singlet pairing is discussed in detail in Ref.~\onlinecite{Kaladzhyan2016}.
The vector  $\dd({\bm k})$  parametrises the odd-parity triplet pairing term. We consider two cases: a TRS in-plane $\dd$ vector, $\dd_\parallel({\bm k}) = (k_y, \,-k_x, \, 0)$, and a TRS breaking out-of-plane $\dd$ vector, $\dd_\perp({\bm k}) = (0, \,0, \, k_x+ik_y)$. The latter case has been introduced to describe the Sr$_2$RuO$_4$ SC \cite{Maeno2003}. We have studied other $\dd$ vectors giving rise to unitary states and found out that these two choices are generic enough to describe 2D anomalous SCs. For the $\dd_\parallel$ case, we have checked that the total angular momentum operator  $M_\parallel^z=L_z+\sigma_z/2$ commutes with the Hamiltonian $\mathcal{H}_0 $ while for the $\dd_\perp$ case, the operators $M_\perp^z=L_z-\tau_z/2$ and $\sigma_z$ commutes with $\mathcal{H}_0$.
Here, ${\bm L}={\bm r}\times{\bm k}$ denotes the orbital momentum operator and $\tau_z$ is the Pauli matrix acting in particle-hole subspace.

\subsection{Band structure}
\begin{figure}[h!]
\includegraphics*[width=\columnwidth]{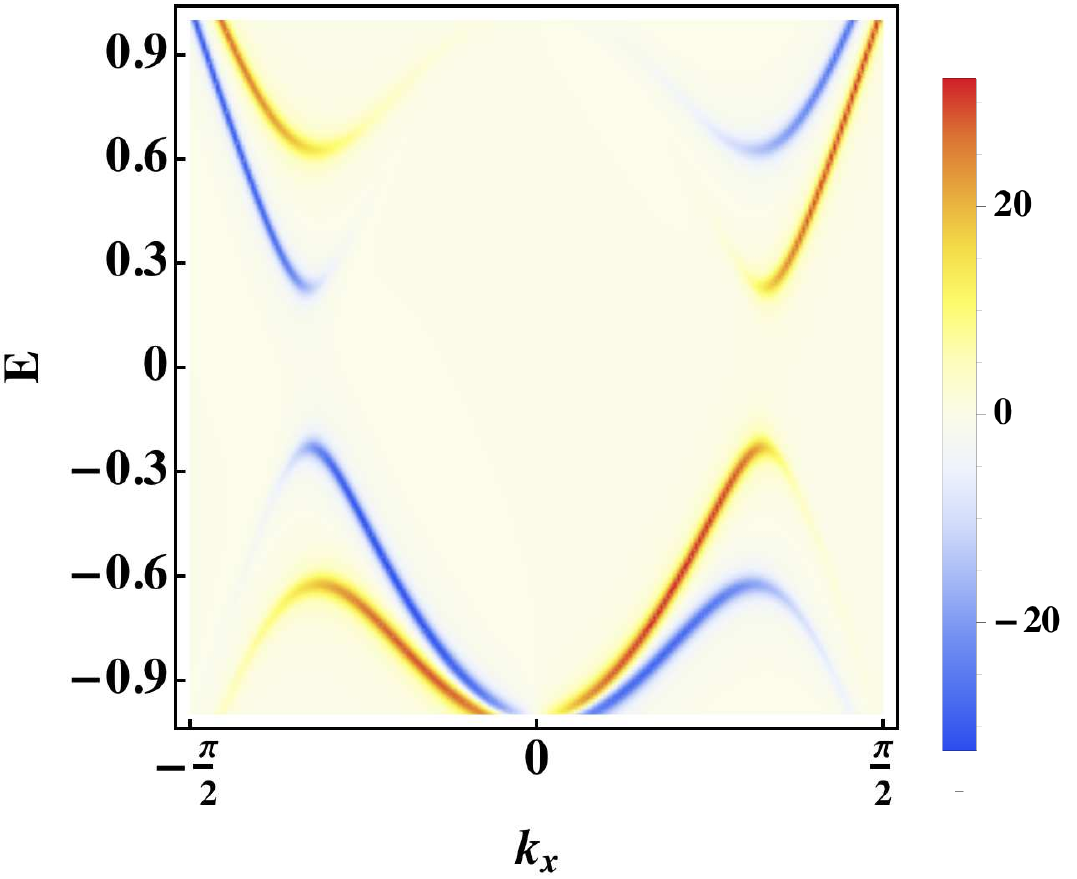}
\caption{Spin-resolved band structure: We plot the $x$ component of the spin-polarized spectral function as a function of energy and $k_x$. We consider a lattice model with dispersion $\Xi_{\bm{k}}=-2t\cos(k_x)-2t\cos(k_y)-\mu$ with $t$ the hopping amplitude and $\mu$ the chemical potential. We take $t=1$, $\mu=-3$,  $\Delta_s=0.2$, $\varkappa=0.5$ and an inverse quasiparticle lifetime $\delta=0.03$.}
\label{bandstructure}
\end{figure}

The energy spectrum of the Hamiltonian given by Eq.~(\ref{h0}) is described by
\begin{eqnarray}
\left| \mathcal{E} \right| = \sqrt{\Xi^2_{\bm{k}}+(\Delta_s\pm\varkappa |{\bm k}|)^2},
\label{spspectrum}
\end{eqnarray}
and the effective superconducting gap is given by
\begin{eqnarray}
\Delta_{eff} = \frac{|\Delta_s-\varkappa k_F|}{\sqrt{1+\varkappa^2}}
\end{eqnarray}
We plot in Fig.~\ref{bandstructure} the spin-resolved band structure (the spin-polarized spectral function) as a function of energy and momentum for the model in Eq. (\ref{h0}) for an in-plane $\mathbf{d}$ vector, $\mathbf{d}=\mathbf{d}_\parallel$. We note that the bands acquire opposite spin polarizations reflecting the helical nature of the superconductor. Note also that the gaps in the two bands are different due to the presence of both s-wave and p-wave couplings. If the s-wave and p-wave coupling become equal, one of the gaps is closing, and the system becomes gapless \cite{Sato2009,Burset2014}: this point marks the transition between an s-dominant and p-dominant regime. 



\subsection{Shiba bound states}
In what follows we study what happens when a  point-like magnetic impurity is introduced in the anomalous SC (see Fig. \ref{Sketch} for a sketch of a magnetic impurity immersed in a p-wave SC with an in-plane $\dd$ vector). The impurities we consider have both a scalar component $U$ and a magnetic component ${\bm J} = (J_x,J_y,J_z)$, and are described by the following Hamiltonian:
\be
\mathcal{H}_{imp} = V \delta(\bm{r}) \equiv
	\begin{pmatrix}
		U \sigma_0 + \bm{J} \cdot \bm{\sigma} & 0 \\
		0 & -U \sigma_0 + \bm{J} \cdot \bm{\sigma}
	\end{pmatrix} \delta({\bm r}),
\ee
where $U$ and ${\bm J}$ are the scalar and magnetic strengths respectively. We assume that only the $l=0$ angular momentum channel matters in our model of the impurity potential. We only consider here classical impurities oriented either along the z-axis, ${\bm J} = (0,0,J_z)$, or along the x-axis, ${\bm J} = (J_x,0,0)$.
%
The spectrum of the Shiba states can be found analytically as detailed in the Appendix A. We generically found two SBS for a point-like impurity. The SBS energies for a magnetic impurity pointing along the z-axis or x-axis are given by:
\be
|E_\pm|= \frac{-\gamma \beta_\pm^2 \sgn \beta_\pm + \sqrt{1 + \beta_\pm^2(1-\gamma^2)}}{1+\beta_\pm^2}  \Delta_t,
\ee
where $\beta\pm = \frac{\pi \nu (U\pm |{\bm J}|)}{\sqrt{1+\varkappa^2}}$ is the dimensionless impurity strength, $\gamma = \frac{\varkappa}{\sqrt{1+\varkappa^2}}$, and $\Delta_t = \frac{\varkappa k_F}{\sqrt{1+\varkappa^2}}$ is the effective p-wave gap.
For a non-magnetic impurity, the two SBS are degenerate while this degeneracy is lifted for a magnetic impurity (see the Appendix A).

In order to analyze the spin structure of these Shiba states, we  resort to the general T-matrix approximation scheme \cite{Balatsky2006}. 
The unperturbed retarded Green's function for our system is $G_0(E,{\bm k})=\left[(E+i\delta)\mathbb{I}_4-\mathcal{H}_0({\bm k})\right]^{-1}$, where $\delta$ is the inverse quasiparticle lifetime. The T-matrix is given by:
\be
T(E) = \left[\mathbb{I}_4-V \int \frac{d^2\mathbf{k}}{(2\pi)^2} G_0(E,\mathbf{k})  \right]^{-1} V
\ee
The FT of the SP LDOS components, $S_{\hat n}({\bm p},E)$, with $\hat{n}=x,y,z$, as well as the FT of the LDOS, $\delta\rho({\bm p},E)$ are found to be:
\begin{align*}
S_x({\bm p},E) &= -\frac{1}{\,2 \pi i\,} \int \frac{d\mathbf{q}}{(2\pi)^2} [\tilde{g}_{12}(E,\mathbf{q},\mathbf{p})+\tilde{g}_{21}(E,\mathbf{q},\mathbf{p})],\\
S_y({\bm p},E) &= -\frac{1}{\,\,2\pi\,\,} \int \frac{d\mathbf{q}}{(2\pi)^2} [g_{12}(E,\mathbf{q},\mathbf{p})-g_{21}(E,\mathbf{q},\mathbf{p})],\\
S_z({\bm p},E) &= -\frac{1}{\,2 \pi i\,} \int \frac{d\mathbf{q}}{(2\pi)^2} [\tilde{g}_{11}(E,\mathbf{q},\mathbf{p}) - \tilde{g}_{22}(E,\mathbf{q},\mathbf{p})], \\
\delta\rho({\bm p},E) &= -\frac{1}{\,2\pi i\,} \int \frac{d\mathbf{q}}{(2\pi)^2} [\tilde{g}_{11}(E,\mathbf{q},\mathbf{p})+\tilde{g}_{22}(E,\mathbf{q},\mathbf{p})],
%
%
\end{align*}
where 
\begin{align*}
g/\tilde{g}(E,\mathbf{q},\mathbf{p}) &= G_0 (E,\mathbf{q}) T(E) G_0(E, \mathbf{p+q}) \\
&\pm G^*_0(E, \mathbf{p+q}) T^*(E) G^*_0 (E,\mathbf{q})
\end{align*}
and $g_{ij}$, $\tilde{g}_{ij}$ signify the corresponding components of the matrices $g$ and $\tilde{g}$.

\section{Results}
We begin by plotting the spatially averaged LDOS and SP LDOS as functions of energy. Note that when performing the T-matrix calculations we consider a discretized version of the Hamiltonian in Eq.~(\ref{h0}) and we perform all the momentum integrals over the first Brillouin zone. The impurity states appear as resonances 
overlapping with a background which corresponds to the DOS in the absence of impurities. Since the unperturbed DOS is not spin-polarized, this background is non-zero only for the averaged LDOS, and the SP DOS contains only impurity-induced contributions. The average DOS and SP DOS are given respectively by 
\begin{align}
\rho(E)&=\rho_0(E)+{\cal N} \delta \rho({\bm p}=0, E) \\
S_{\hat n}(E)&=S_{\hat n}({\bm p}=0,E),
\end{align}
for $\hat n=x,y,z$. Here ${\cal N}$ denotes the impurity concentration.
In this paper, we consider the dilute impurity limit and thus we take in what follows a concentration of impurities of ${\cal N}=2\%$. 

\begin{figure}[h!]
	\includegraphics*[width=0.48\columnwidth]{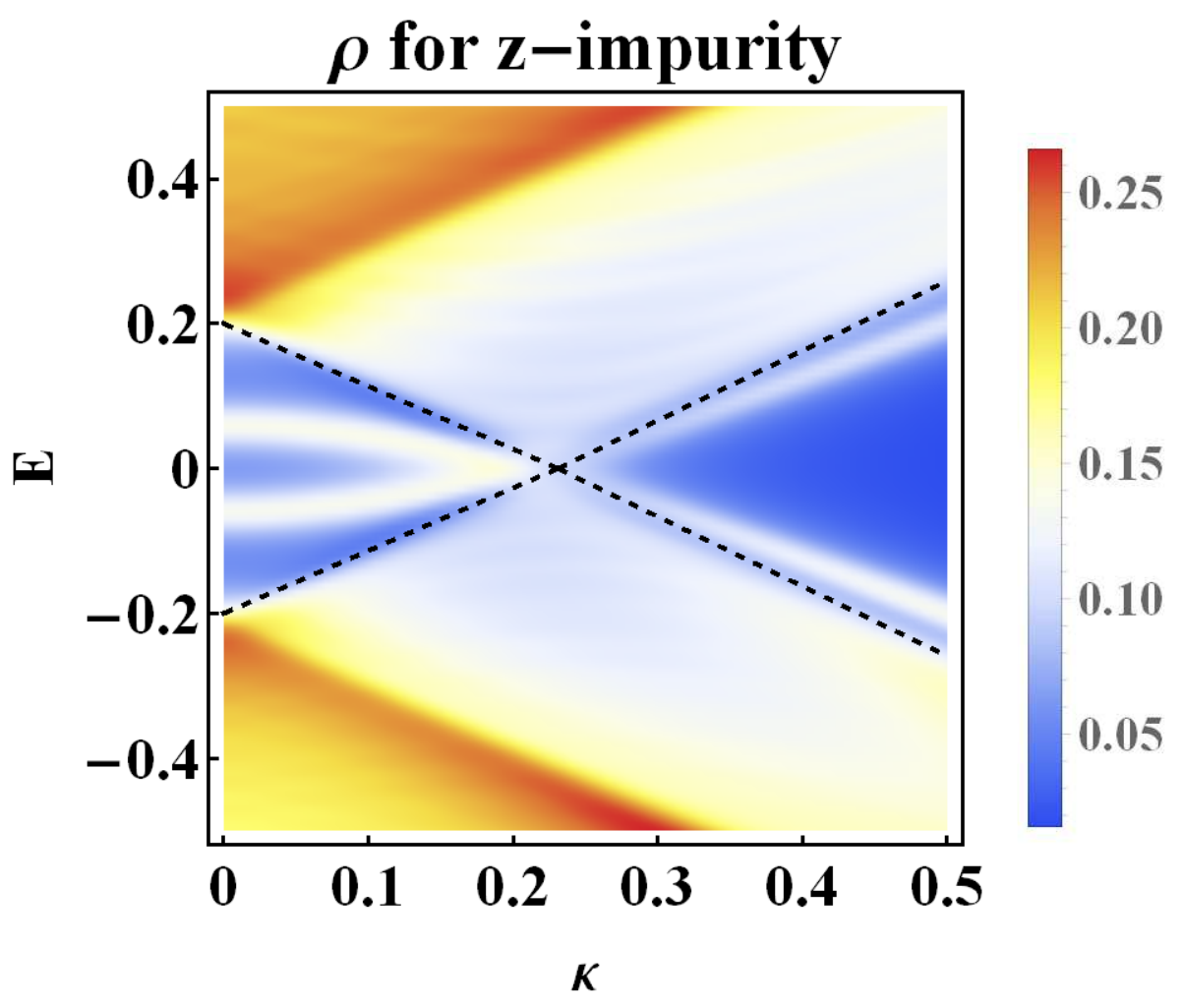}
	\includegraphics*[width=0.46\columnwidth]{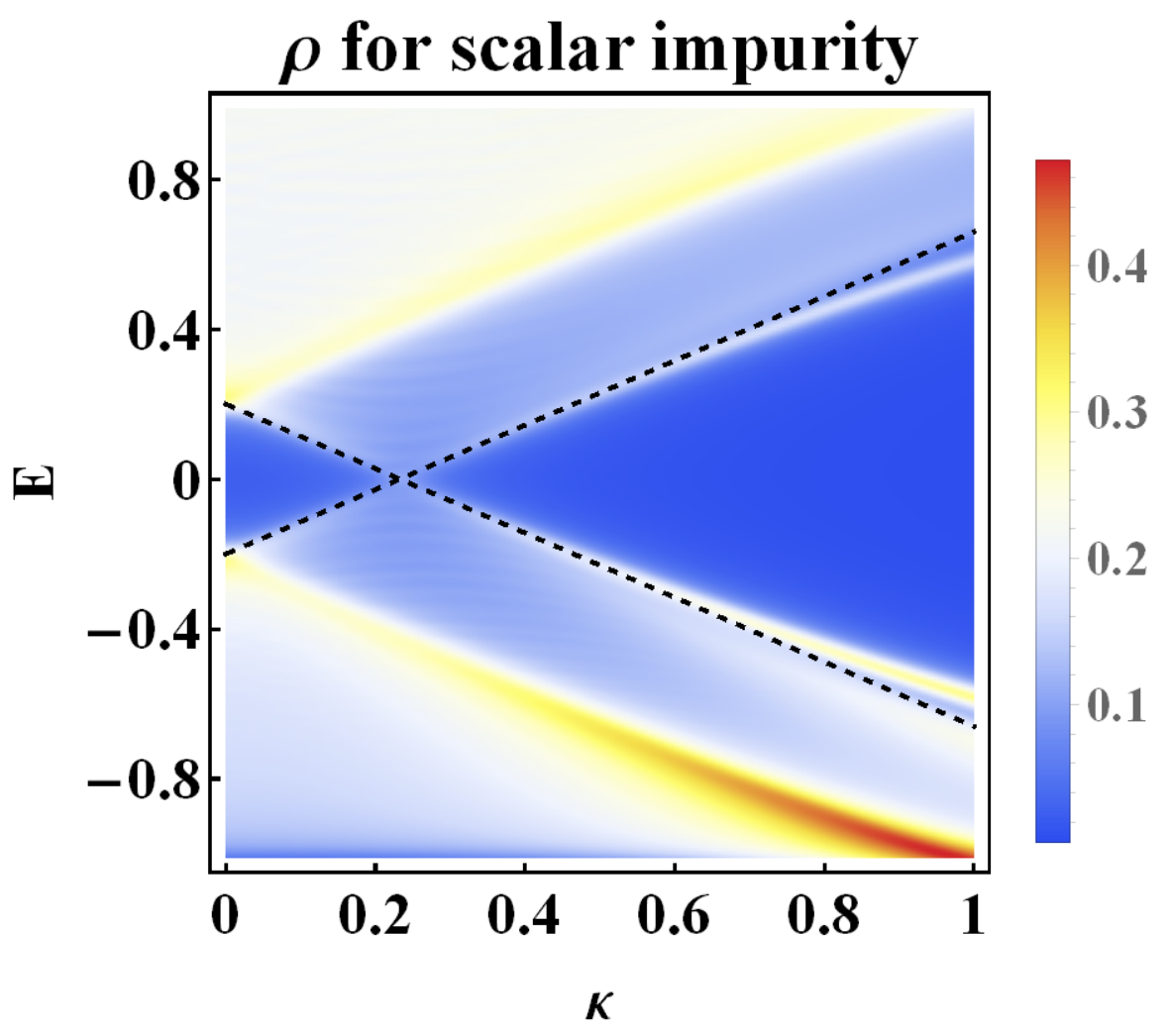}\\
	\includegraphics*[width=0.48\columnwidth]{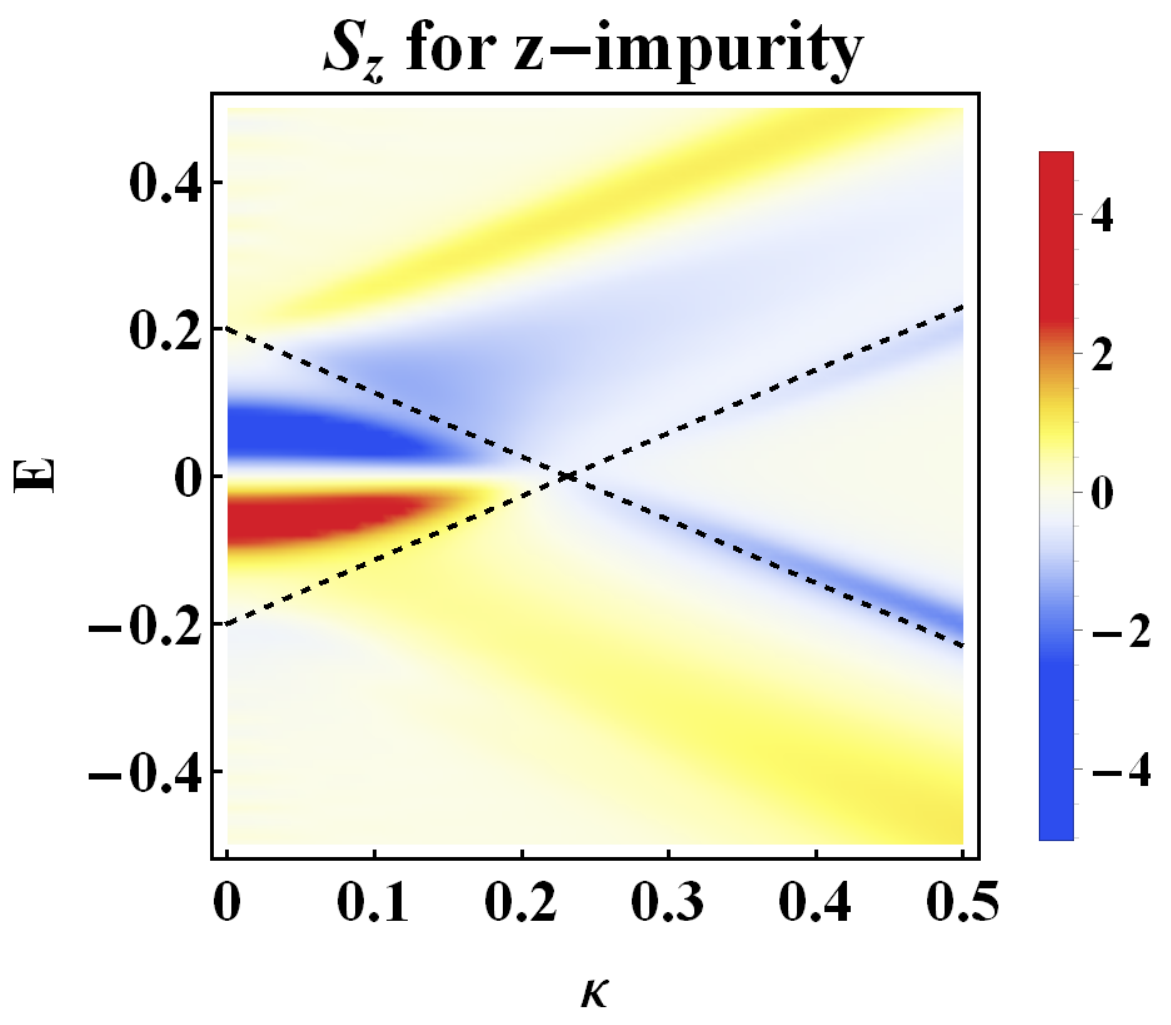}	
	\includegraphics*[width=0.48\columnwidth]{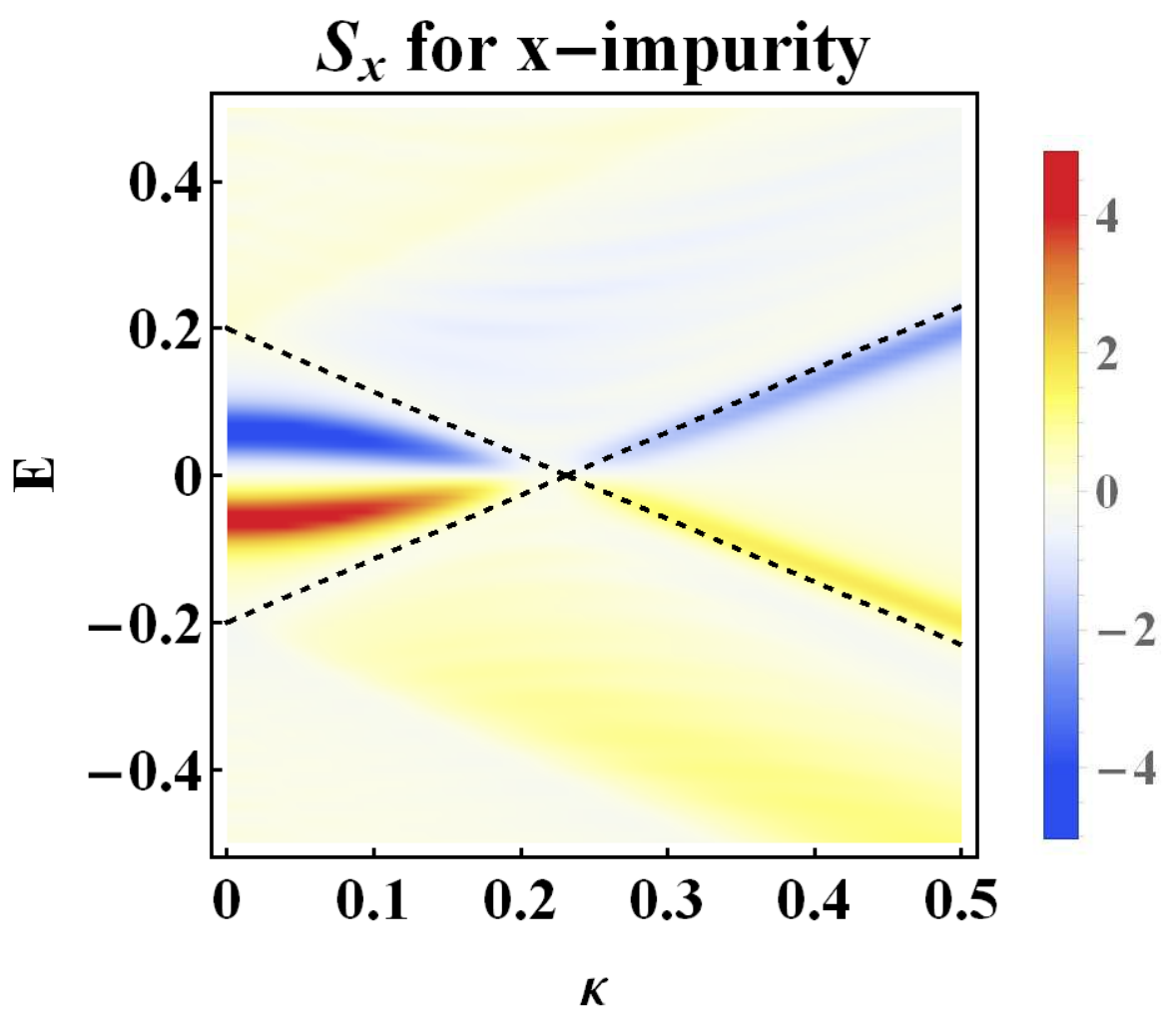}
	\caption{(Color online) Average LDOS (first row) and SP LDOS (second row) (in arbitrary units) as functions of energy and the p-wave pairing, $\varkappa$, for an in-plane $\dd$ vector.  We consider a scalar impurity with 
$U=6$ in the upper right panel, and magnetic impurities with an impurity strength of $J_z=2$ (left column) and with $J_x=2$ (lower right panel). We set $\Delta_s=0.2$, and $\delta=0.03$. The gap of the system is denoted by the dashed line.}
	\label{kappavar}
\end{figure}

\begin{figure}[h!]
	\includegraphics*[width=0.48\columnwidth]{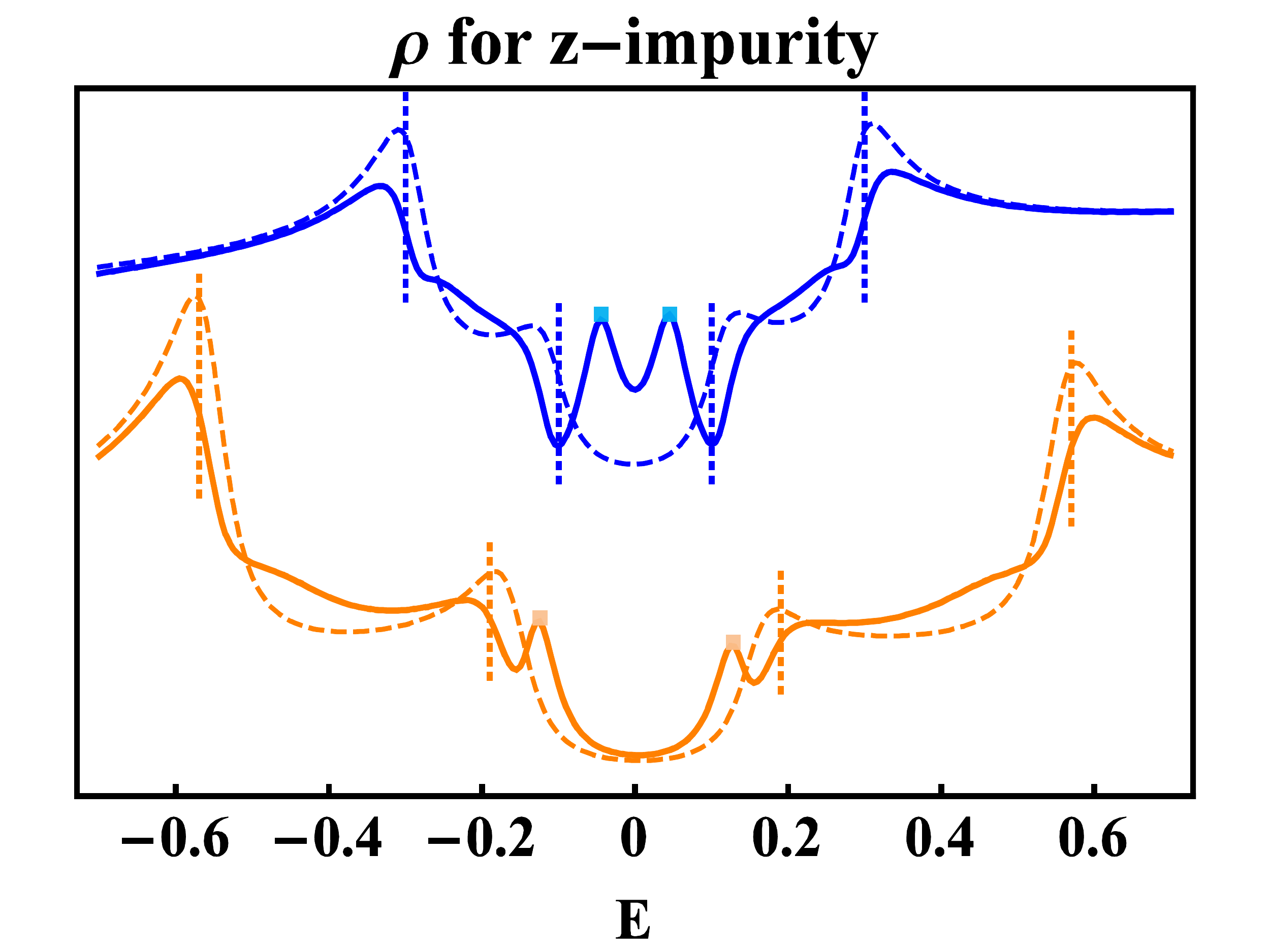}
	\includegraphics*[width=0.48\columnwidth]{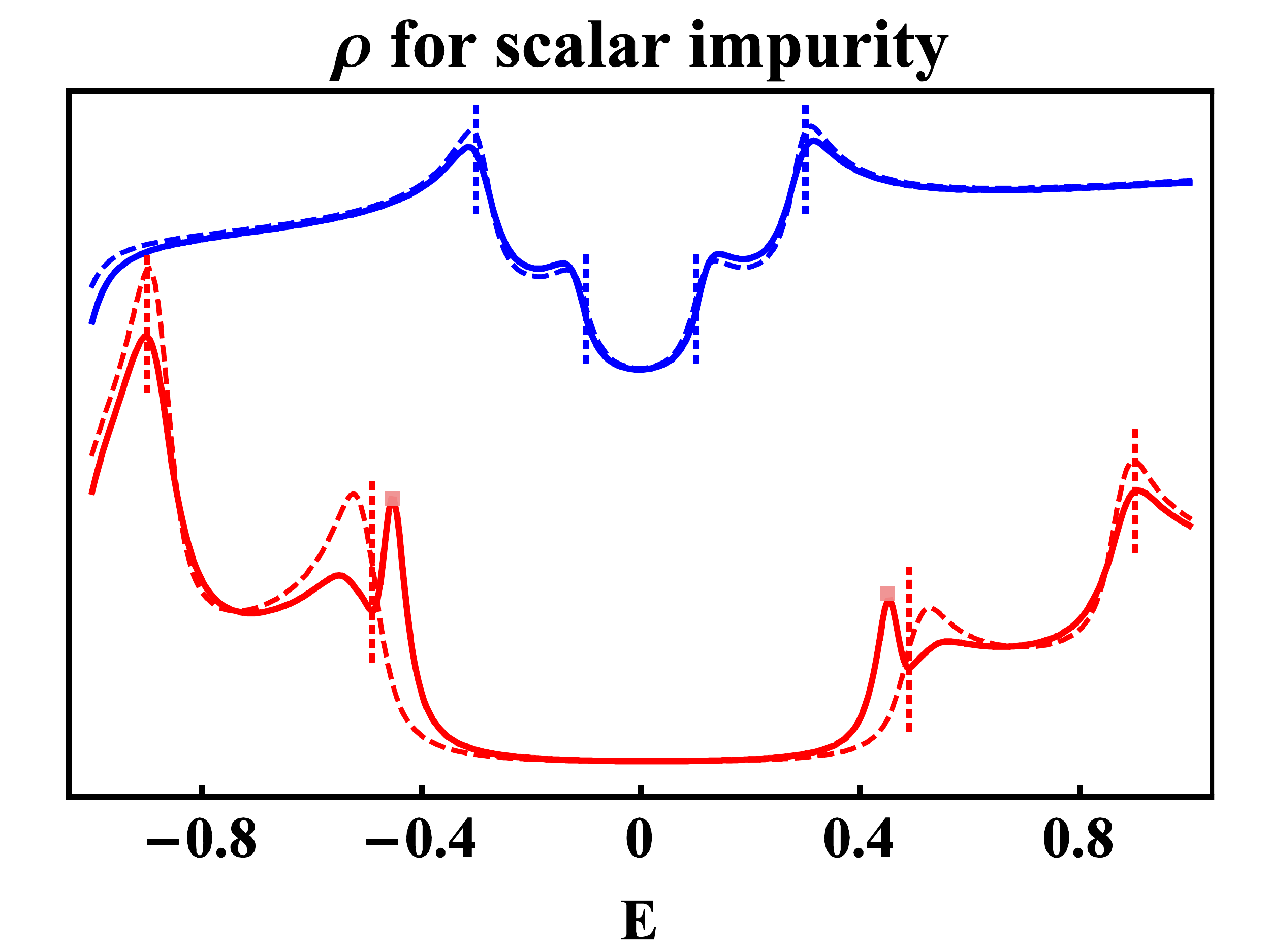}\\
	\includegraphics*[width=0.48\columnwidth]{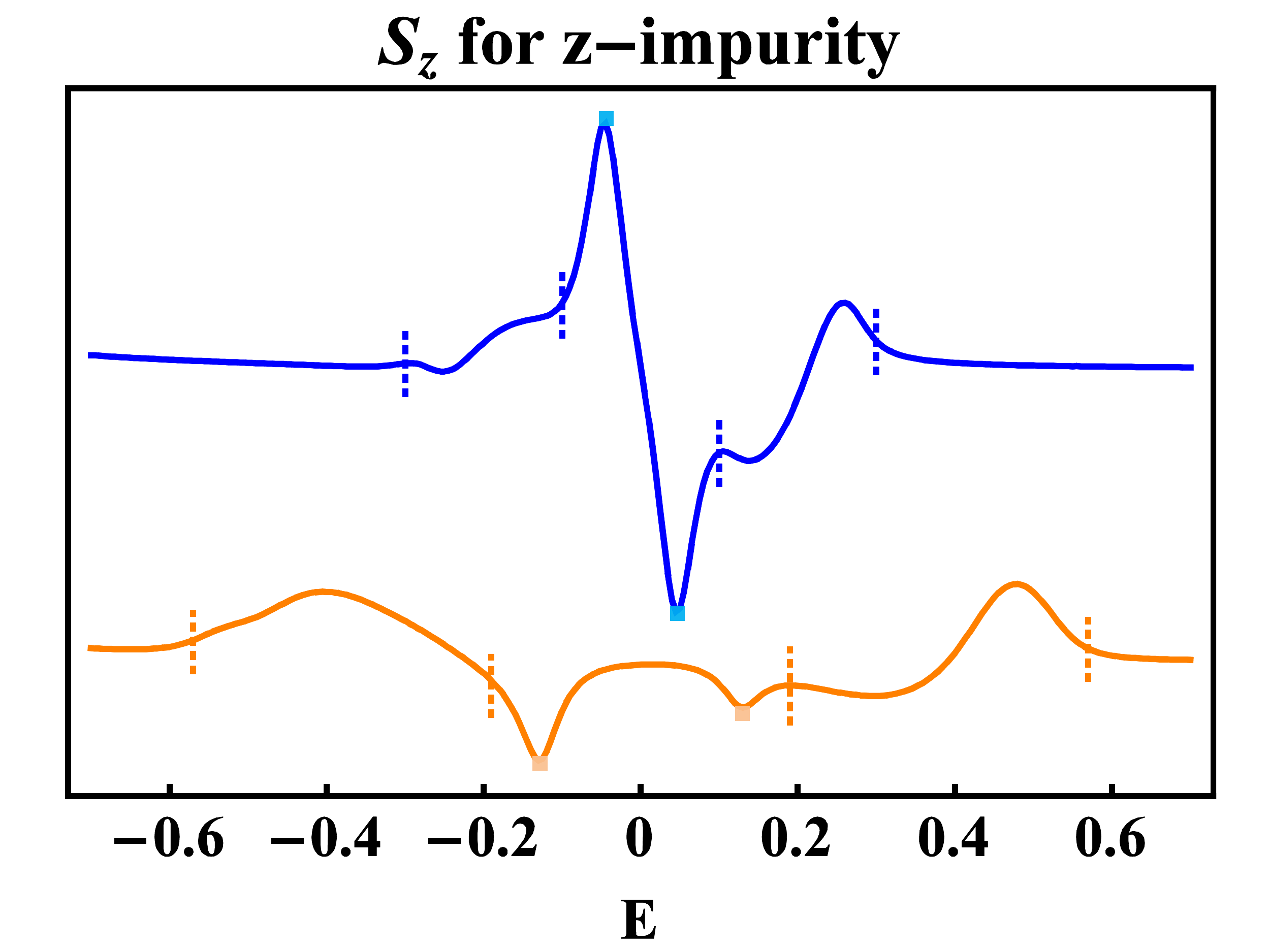}	
	\includegraphics*[width=0.48\columnwidth]{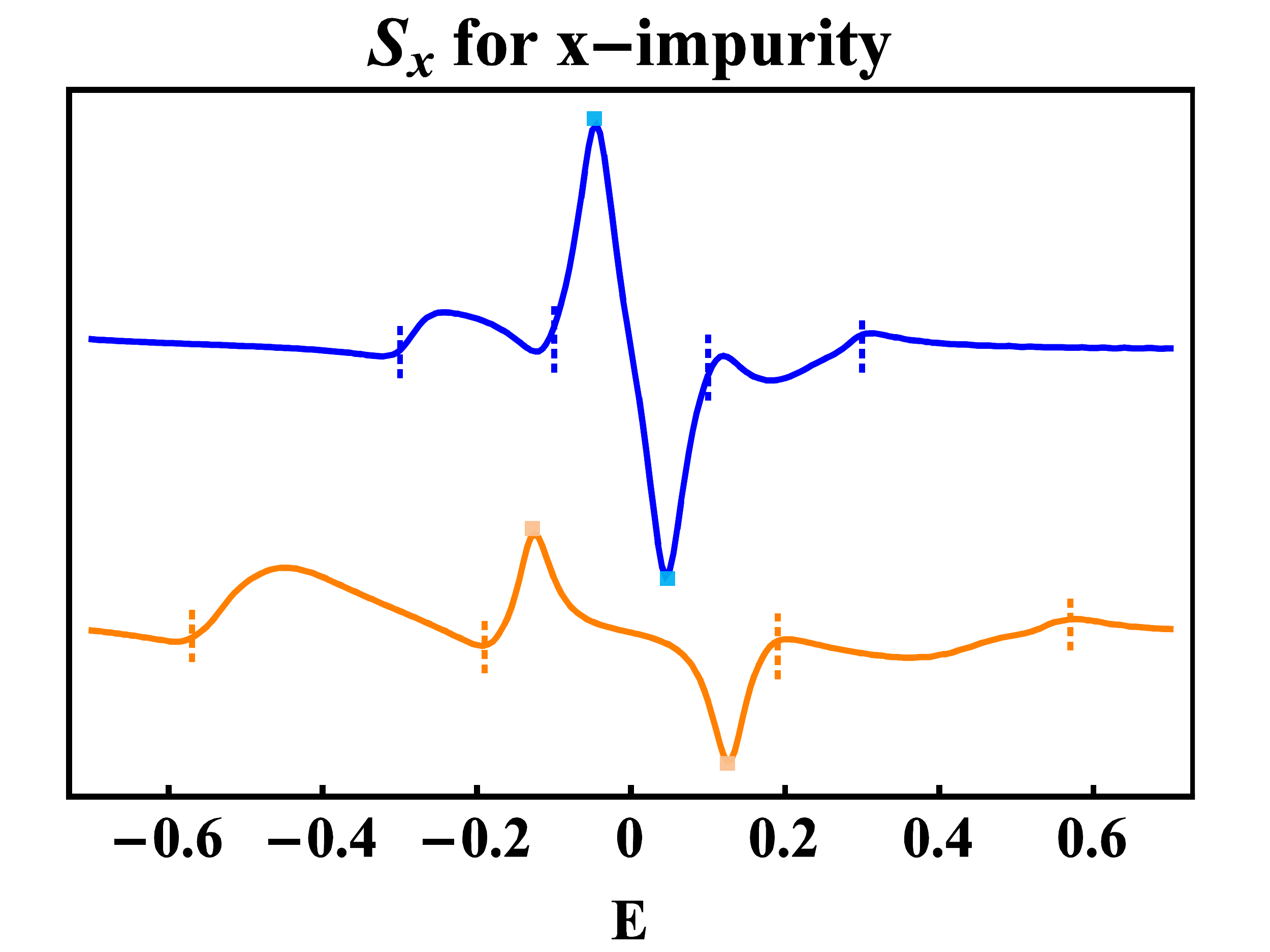}
	\caption{(Color online) 
Upper panel:
	Average LDOS (in arbitrary units) as functions of energy  for an in-plane $\dd$ vector for different values of the p-wave pairing, $\varkappa$: blue stands for $\varkappa = 0.1$, orange for $\varkappa=0.4$ and red for $\varkappa=0.8$.
The dashed lines correspond to the averaged LDOS in the absence of impurities while the plain lines show the LDOS in presence of impurities. We can thus disentangle the spectral features due to the impurity bound states from the gap edges.
Lower panel: same as the upper panel but for the SP LDOS. The vertical dashed lines correspond to the two gap edges denoted in the main text by $\Delta_{eff}^{\pm}$. The subgap Shiba states peaks are marked by filled squares. The other parameters are the same as in Fig. \ref{kappavar}.
}
	\label{cuts}
\end{figure}

In Figs.~\ref{kappavar}, \ref{cuts}, \ref{spJvar} and  \ref{purepJvar}  we focus on how the impurity states are affected by the value of $\varkappa$, the p-wave order parameter, as well as by the impurity strength ${\bm J}$. Note that an impurity with magnetic moment 
along a direction specified by a unit vector $\hat{n}$, gives rise to a total  (averaged along the entire space) non-zero polarization {\em only} along  $\hat{n}$, even if it would give rise to a non-zero spatial spin structure in more than one spin components. 


\subsection{Number of Shiba bound states}
\begin{figure}
	\centering
	\begin{tabular}{ccc}
		  & $S_z \phantom{lo}$ & $S_x \phantom{lo}$ \\
		\rotatebox{90}{$\phantom{loremipsun}$E} &  \includegraphics*[width=0.48\columnwidth]{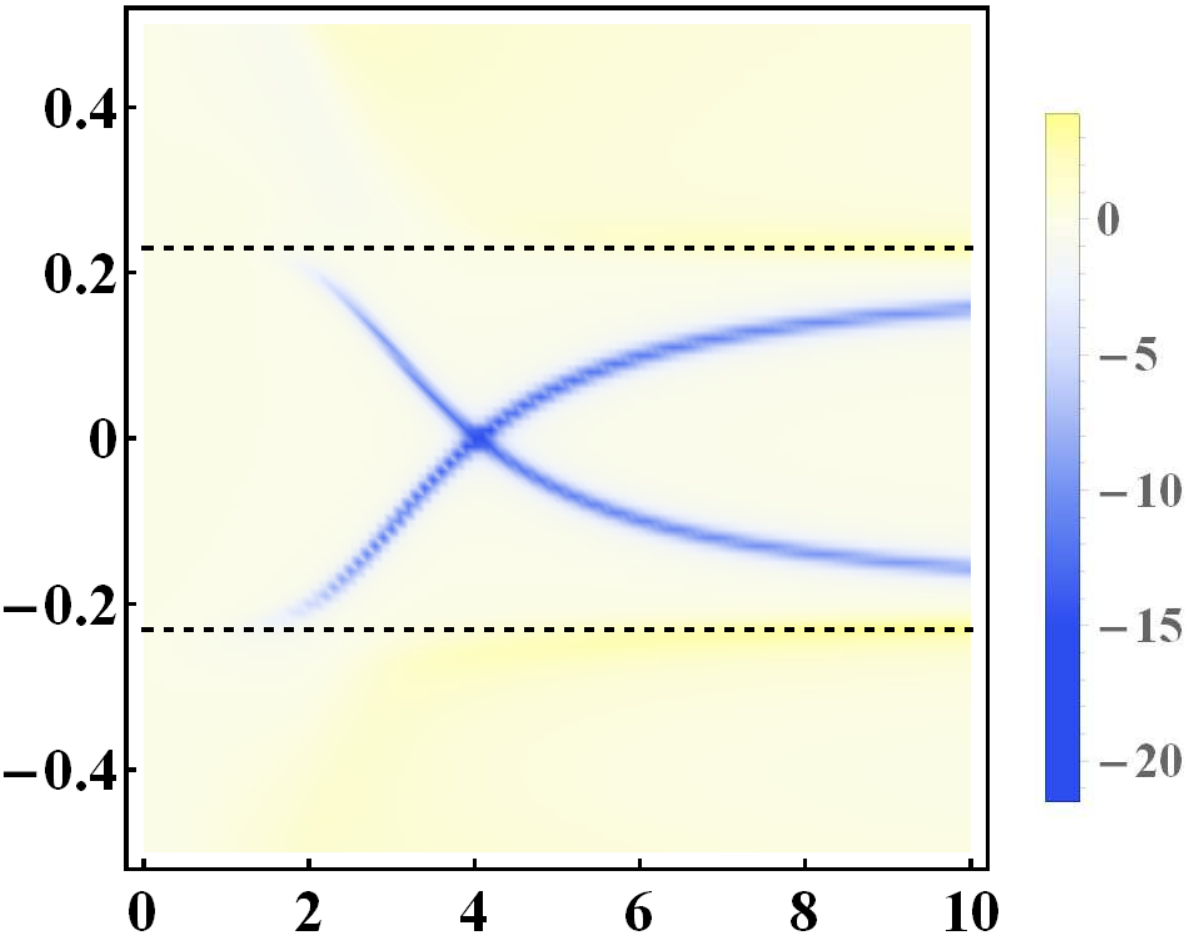}  &  \includegraphics*[width=0.48\columnwidth]{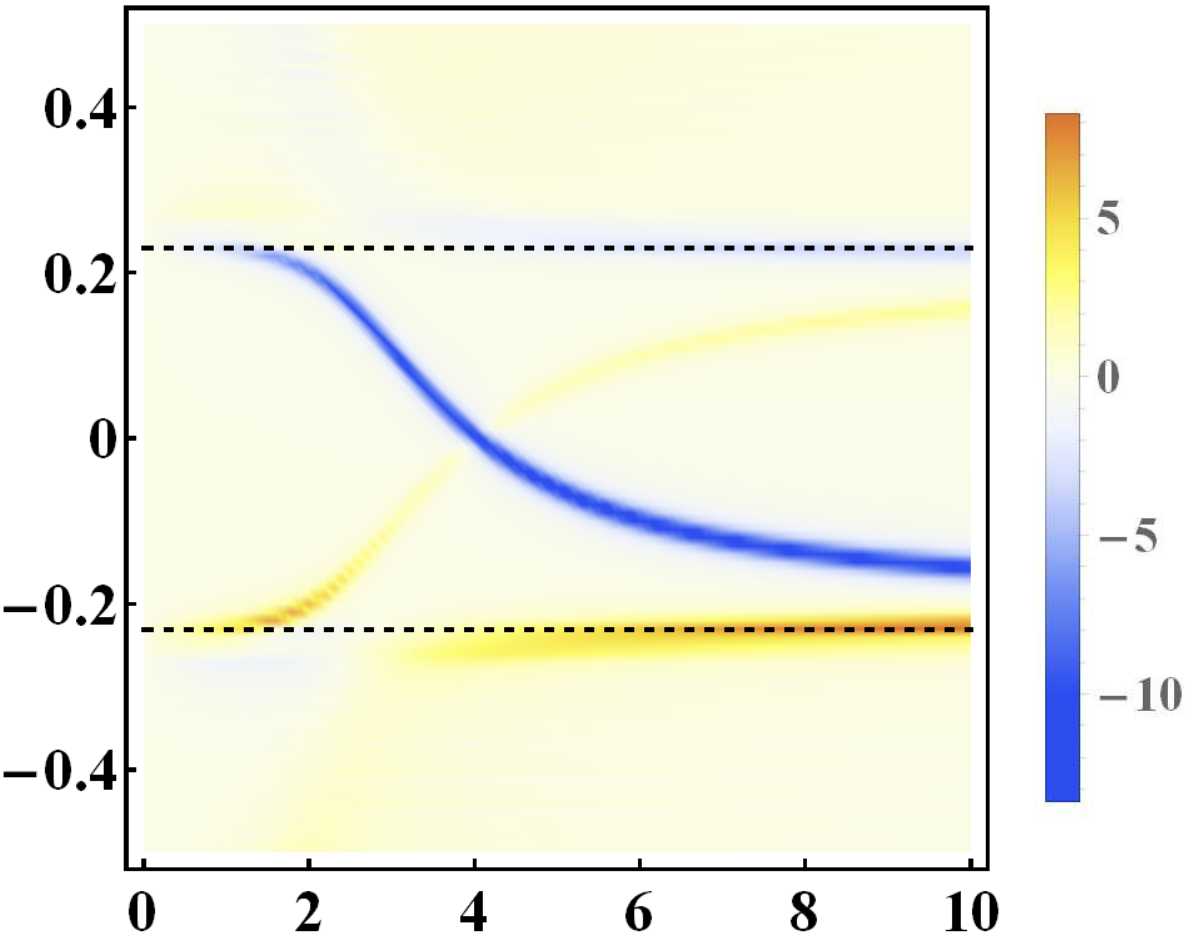}  \\
		\rotatebox{90}{$\phantom{loremipsun}$E} &  \includegraphics*[width=0.48\columnwidth]{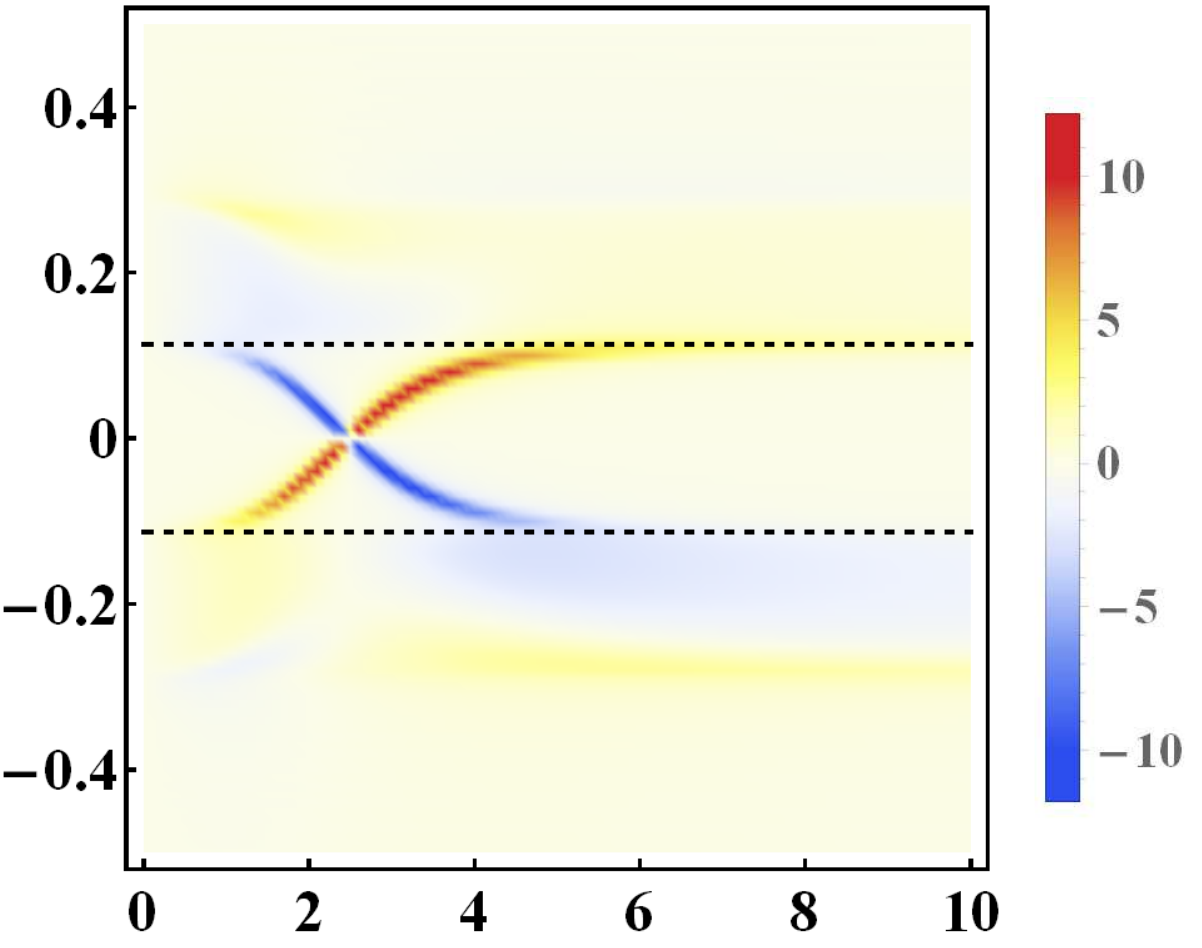}   &  \includegraphics*[width=0.48\columnwidth]{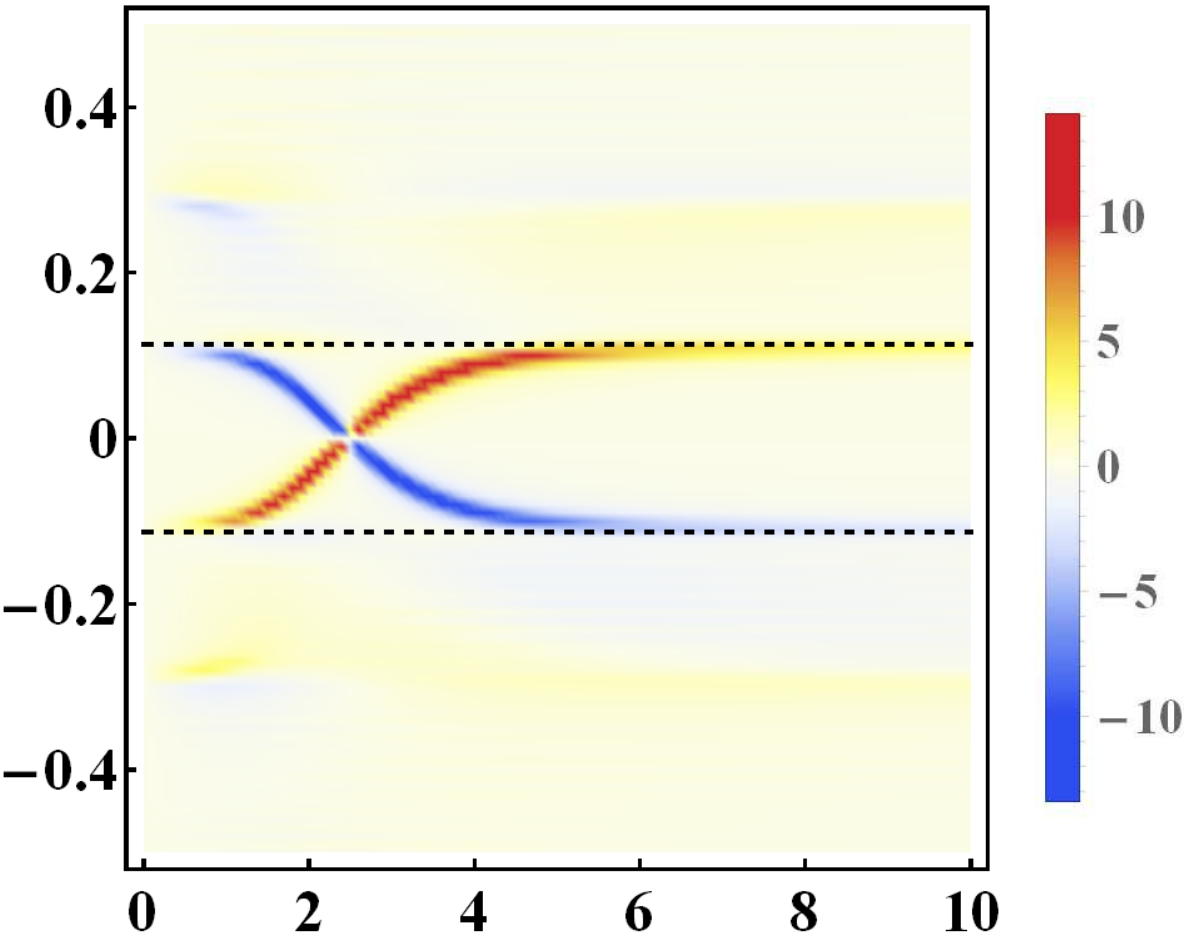}  \\
  		& $J_z \;(J_x = J_y = 0)\phantom{aa}$ & $J_x \;(J_y = J_z = 0)\phantom{aa}$
\end{tabular}
	\caption{(Color online) Average SP LDOS as a function of energy and impurity strength for magnetic impurities with spin along $z$ (left column) and $x$ (right column) for an in-plane $\dd$ vector. We set $\delta=0.01$ and we focus on the dominant p-wave case ($\varkappa=0.5$, $\Delta_s=0.2$, first row), and the dominant s-wave case ($\varkappa=0.1$, $\Delta_s=0.2$, second row). The gap is denoted by the dashed line.}
		\label{spJvar}
\end{figure}

\begin{figure}
	\centering
	\begin{tabular}{cccc}
		 & & $S_z \phantom{lo}$ & $S_x \phantom{lo}$ \\
		\rotatebox{90}{$\phantom{lorem}$in-plane $\dd_\parallel$} & \rotatebox{90}{$\phantom{loremipmi}$E} &  \includegraphics*[width=0.45\columnwidth]{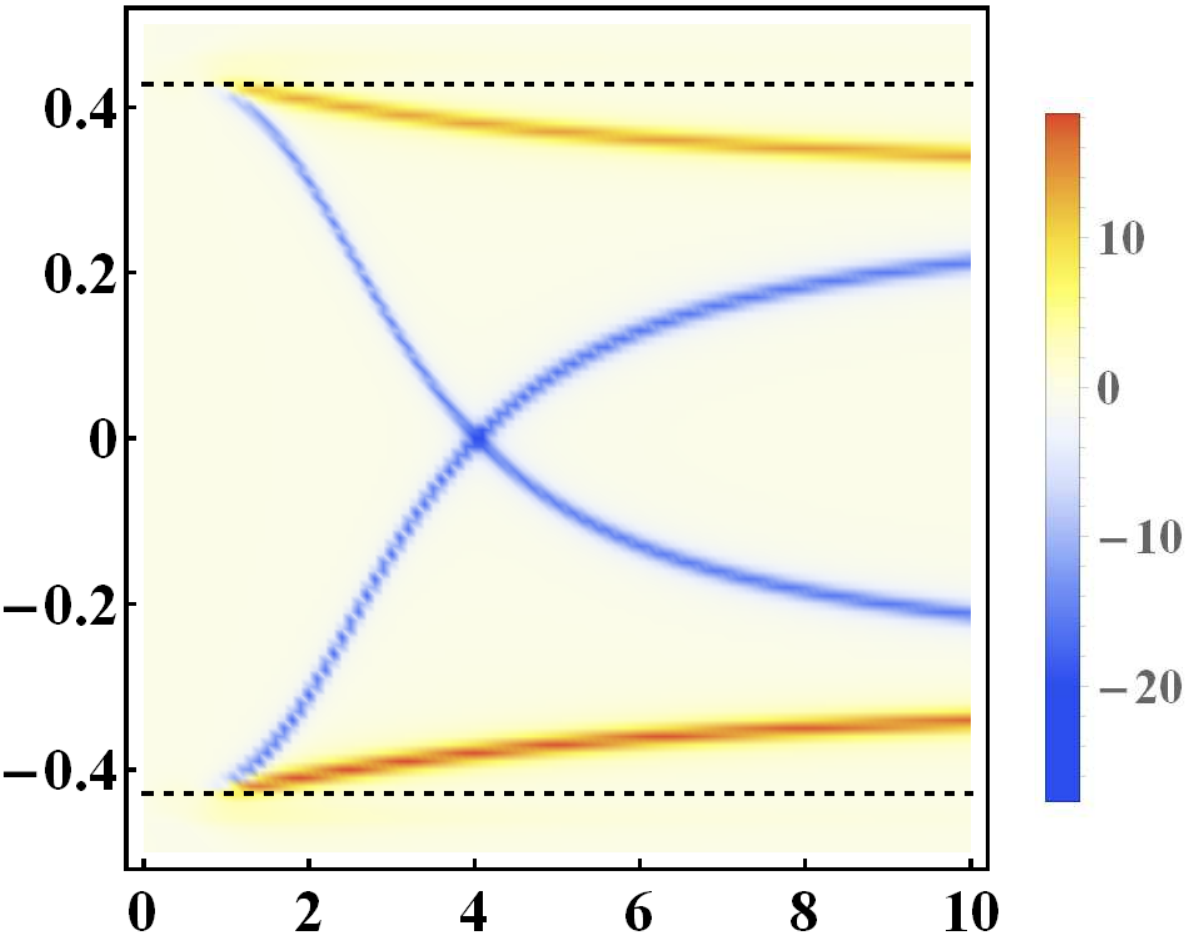}  & \includegraphics*[width=0.45\columnwidth]{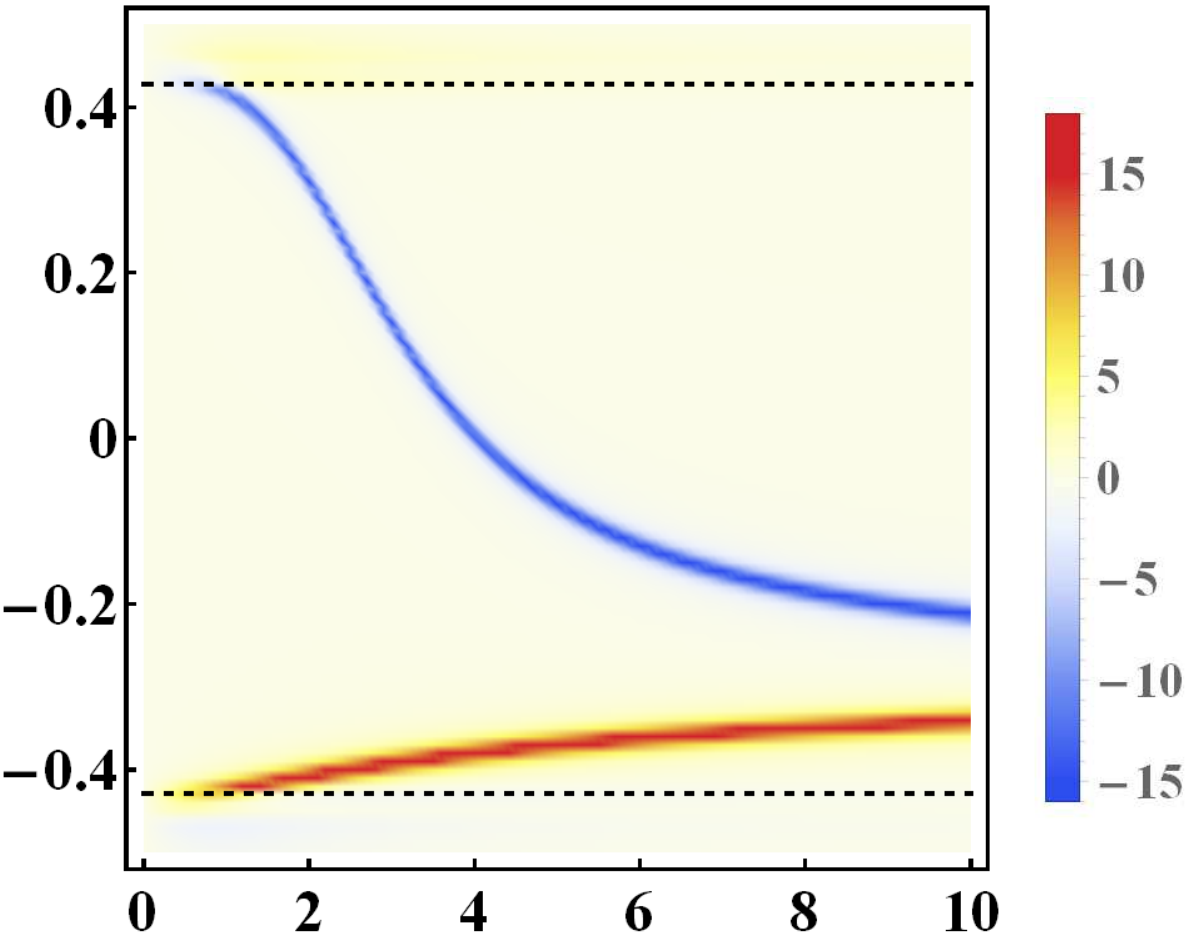}   \\
		\rotatebox{90}{$\phantom{lore}$out-of-plane $\dd_\perp$} & \rotatebox{90}{$\phantom{loremipmi}$E} &  \includegraphics*[width=0.45\columnwidth]{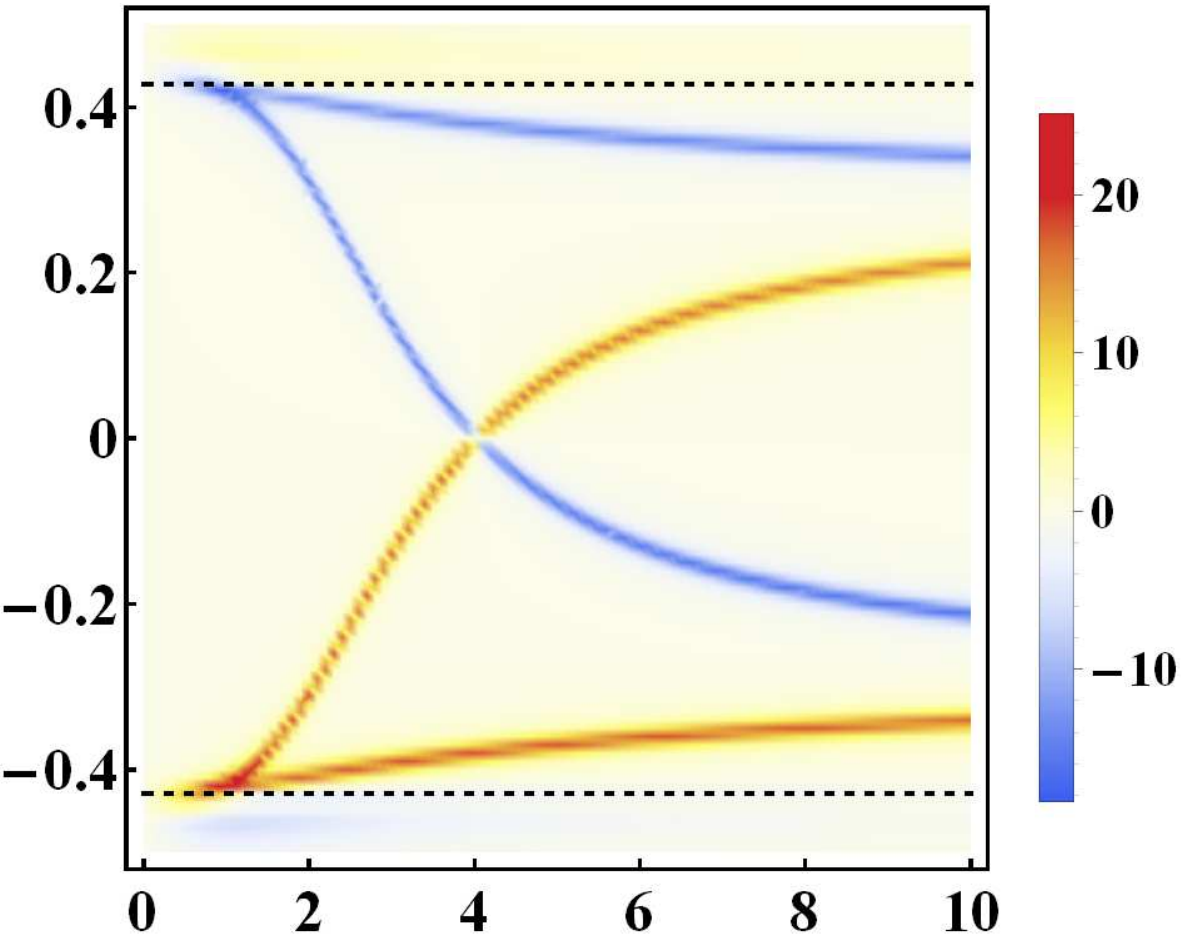}  & \includegraphics*[width=0.45\columnwidth]{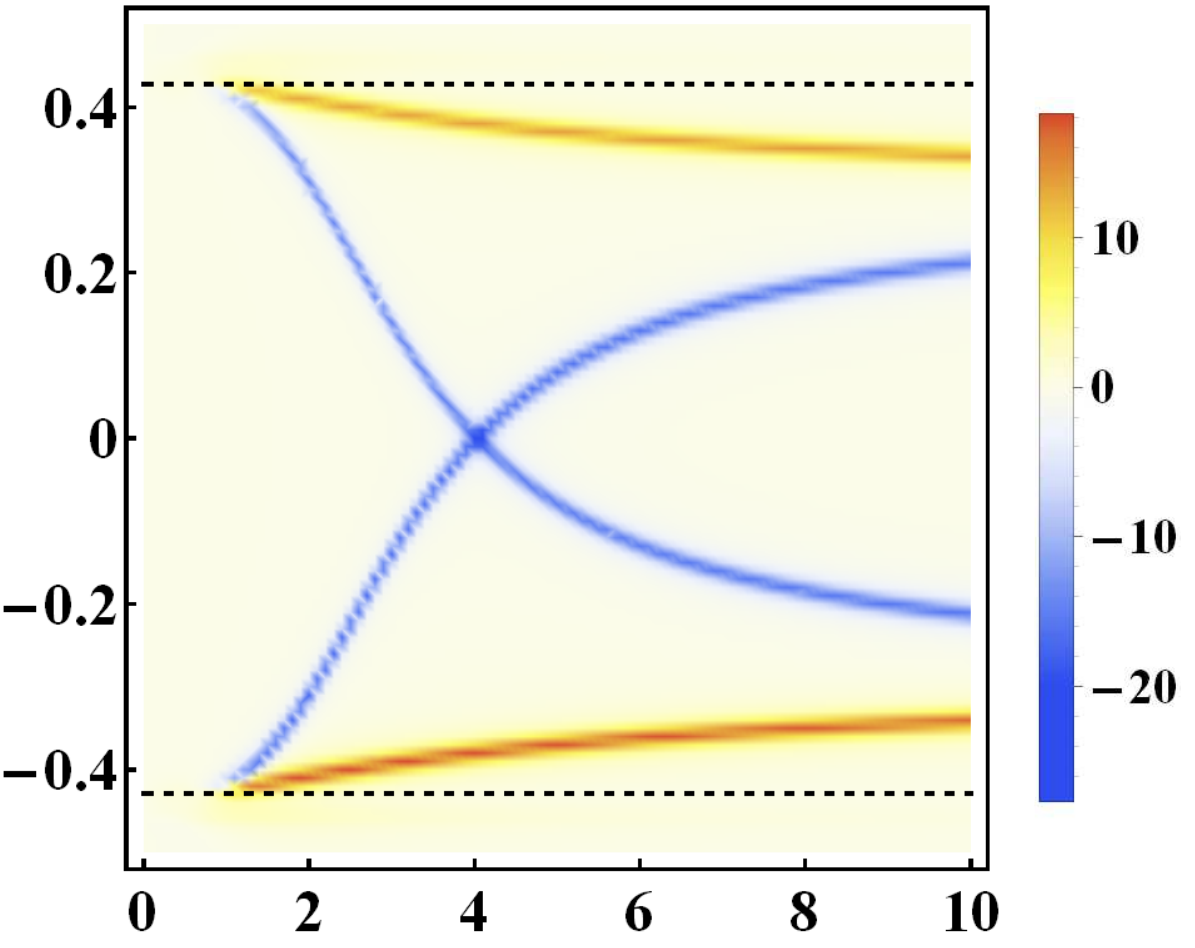}   \\
  		& & $J_z \;(J_x = J_y = 0)\phantom{aa}$ & $J_x \;(J_y = J_z = 0)\phantom{aa}$
\end{tabular}
	\caption{(Color online) Average SP LDOS as a function of energy and impurity strength for magnetic impurities with spin along $z$ (left column) and along $x$ (right column). We take $\delta=0.01$ and we focus on the pure p-wave $\varkappa=0.5$, $\Delta_s=0$ with in-plane $\dd_\parallel$ (first row) and out-of-plane $\dd_\perp$ (second row). The gap is denoted by the dashed line.}
	\label{purepJvar}
\end{figure}

The number of Shiba states depends on the type of impurity, as well as on the two competing SC order parameters. Therefore,  there is no sub-gap SBS in the non-topological s-dominant regime for a scalar impurity, and a double-degenerate SBS in the topological p-dominant regime \cite{Balatsky2006,Lutchyn2015} (see also [\onlinecite{Sau2013}] and the Appendix B). As shown in Figs.~\ref{kappavar} and \ref{cuts}, these states tend to stay close to the gap edge for not too large p-wave couplings. For a magnetic impurity, one SBS forms in the s-dominant regime, and two in the p-dominant regime (see Figs.~\ref{kappavar}, \ref{cuts} and \ref{spJvar}). Out of the two, in most cases one is a subgap state, while the other  is dissolved in the continuum. When the impurity strength increases this  `bulk' SBS  approaches the gap edge and becomes more visible (see Fig.~\ref{purepJvar} and also the SI \cite{sm}).
However, this is not a generic feature. All  impurity states can be subgap states, as it does occur for a pure p-wave SC (see Fig.~\ref{purepJvar}), as well as for a small s-wave coupling (see Fig. S2 in the SI \cite{sm}). When increasing the impurity strength the energy of the  inner BS decreases to zero and there is a level crossing corresponding to a change of  the ground state parity \cite{Sakurai1970,Balatsky2006} (see Figs.~\ref{spJvar} and \ref{purepJvar}). As can be seen from these two figures the pure p-wave and the dominant p-wave regime are very similar, the main difference being the position of the outer SBS with respect to the band edge.

\subsection{Local density of states}
In the upper panel of Fig.~\ref{kappavar}, we plot the average LDOS as a function of the triplet pairing parameter $\varkappa$ and energy. This allows us to visualize the sub-gap states together with the edges of the gap. In order to illustrate the spectral features of the averaged LDOS, we plot
in the upper panel of Fig.~\ref{cuts}  vertical cuts of the density plots shown in Fig.~\ref{kappavar}. The dashed lines mark the unperturbed averaged LDOS. In the presence of both s-wave and p-wave pairing, there exist two effective values giving the gap edges (for derivation see the Appendix C): 
\begin{equation}
\Delta^\pm_{eff} = \frac{|\Delta_s \pm \varkappa k_F|}{\sqrt{1+\varkappa^2}}.
\end{equation}
It is clear that some of the noticeable out-of-the-gap features appearing in Fig.~\ref{kappavar} can be traced back to the gap edges (denoted by the dashed lines) slightly modified by the presence of the impurity that affects all the energies in the continuum. Besides we can identify features corresponding to the localized impurity, which we identify by filled squares. The subgap Shiba states peaks are identified by filled triangles. Note that the asymmetry in the averaged LDOS between positive and negative energy $E$, 
is a direct consequence of the fact that  we plot only the electronic parts of the Green's function.


\subsection{Spin-polarized local density of states}

As a first striking result, we find that for the `orthogonal' configurations (either an in-plane $\dd$ vector  and an out-of-plane impurity spin, or  the reverse situation), both particle and hole components of the  SBSs closest to midgap have electronic spins with the same spin orientation
in the p-dominant regime, and of opposite orientation in the s-dominant regime. This is also shown in the upper left panel of Figs. \ref{spJvar} and \ref{purepJvar} for the former case (see also Fig. S3 in the SI \cite{sm} for the latter case). 
This is a direct signature of the triplet nature of the Cooper pair and of  the $\dd$ vector orientation. Indeed, the spins of the paired electrons live in the plane  orthogonal to the $\dd$ vector (see Fig. \ref{Sketch}). For an in-plane $\dd$ vector, the paired electrons have, therefore, always a non-zero spin component along the z direction. Because of the antiferromagnetic exchange coupling, it thus costs less energy to break Cooper pairs with spins pointing in the direction opposite to the impurity spin. Note that in the latter case $M^z_\parallel$ commutes with the full Hamitlonian $\mathcal{H}_{0} +\mathcal{H}_{imp}$ which allows to attribute a well defined angular momentum $M^z_\parallel=\pm 1/2$ to the Shiba states.
By analogy with the local susceptibility  of triplet SCs \cite{Sigrist2009} this explains the spin sensitivity  of the SBS  to a local impurity spin pointing along  the z axis.

On the other hand, when the $\dd$ vector and the impurity spin are both in-plane (Fig.~\ref{kappavar} lower right panel) or 
both out-of-plane (lower left panel of Fig. \ref{purepJvar}, and  also Fig. S3 in the SI \cite{sm}), the opposite-energy SBS closest to midgap 
have spins of opposite sign for both the s-wave and p-wave dominant regimes  as expected from the previous argument based on the spin orientation of the paired electrons.

\begin{figure}
	\centering
	\begin{tabular}{ccc}
		\multicolumn{3}{c}{$\phantom{o}$\textbf{x-impurity}} \\
		& $S_x \phantom{lo}$ & $S_y \phantom{lo}$ \\
		\rotatebox{90}{$\phantom{lorem}$pure p-wave, $\dd_\parallel$} &   \includegraphics*[width=0.48\columnwidth]{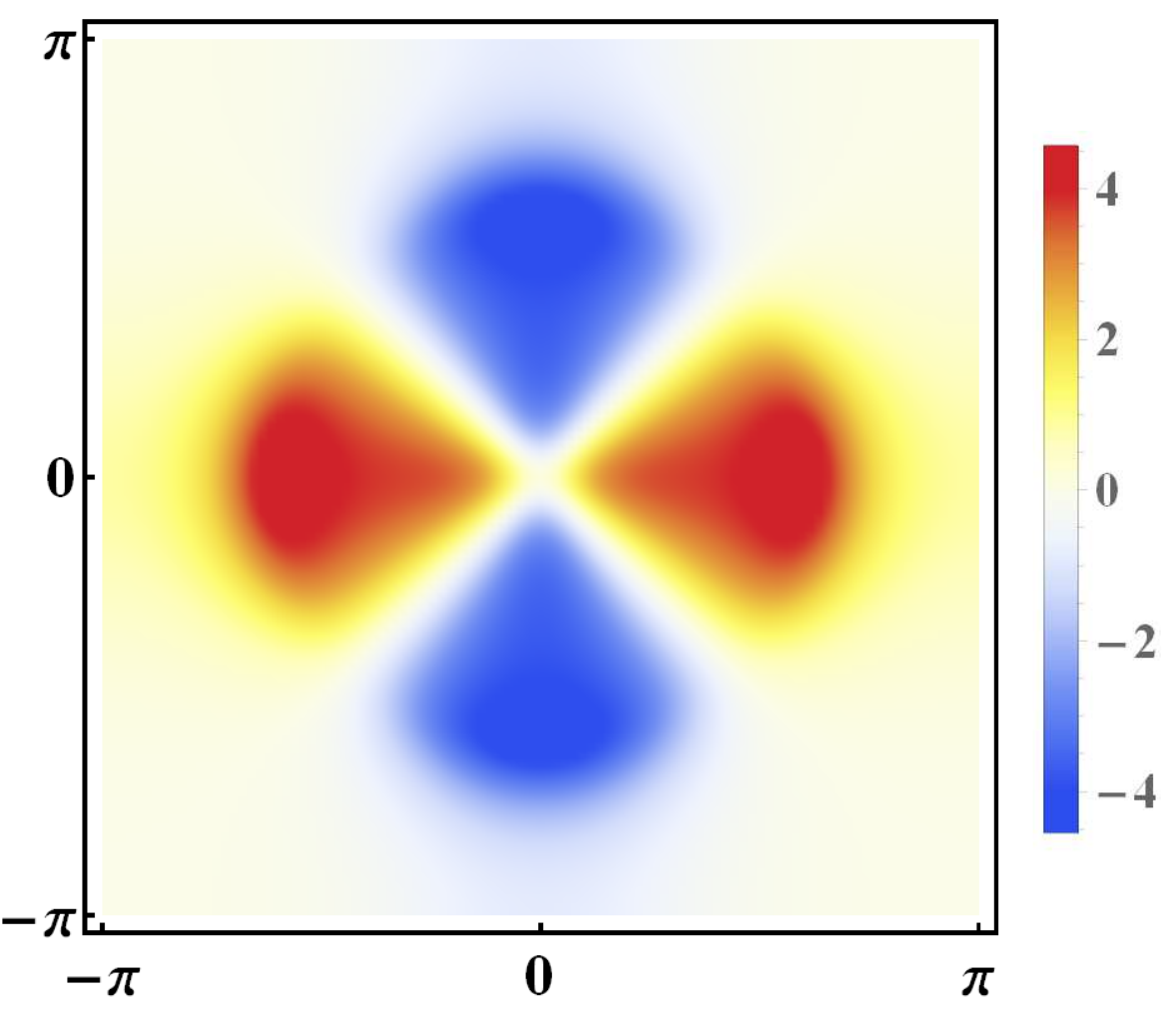}  & \includegraphics*[width=0.48\columnwidth]{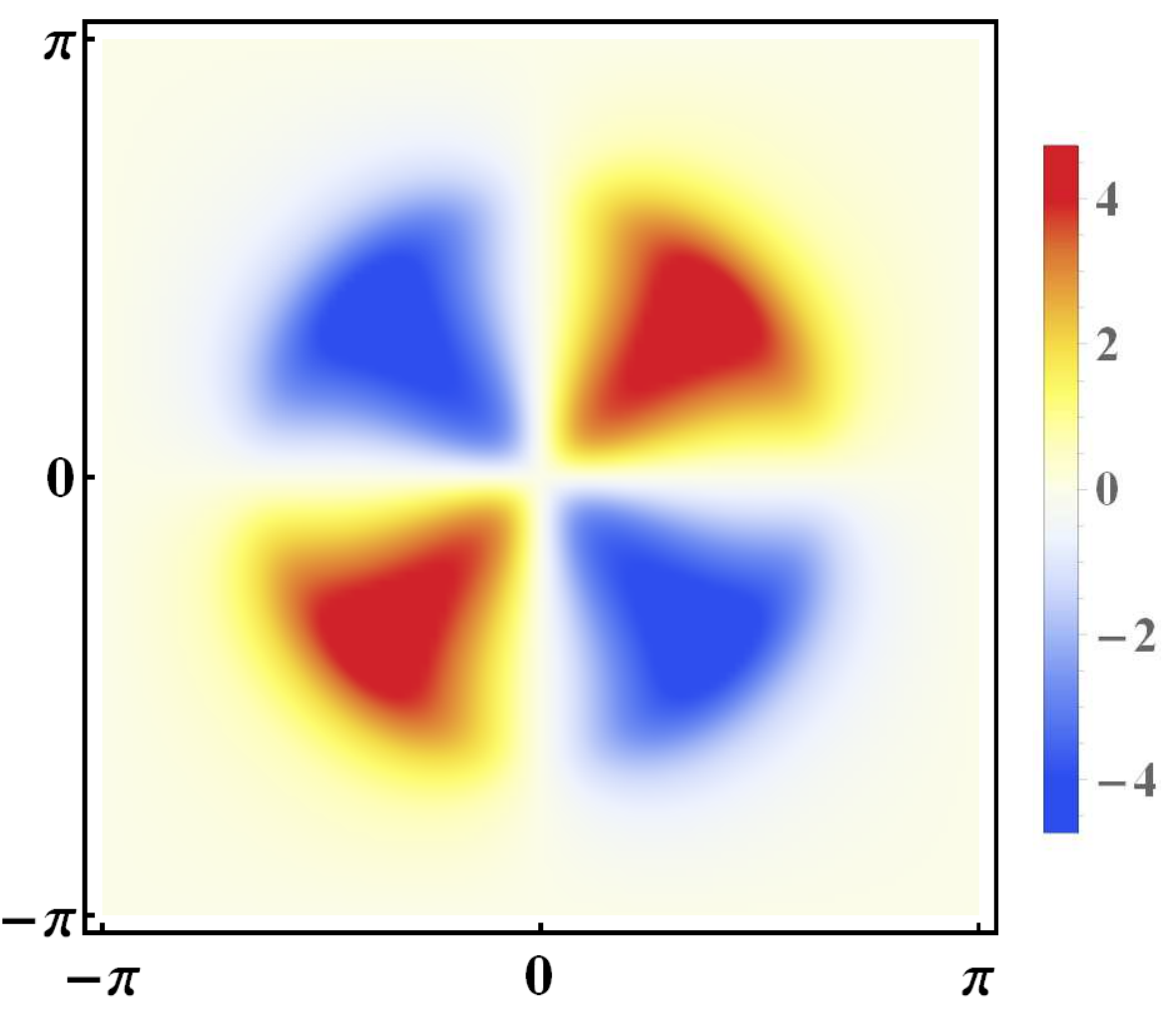}   \\
		\rotatebox{90}{$\phantom{lorem}$pure p-wave, $\dd_\perp$} & \includegraphics*[width=0.48\columnwidth]{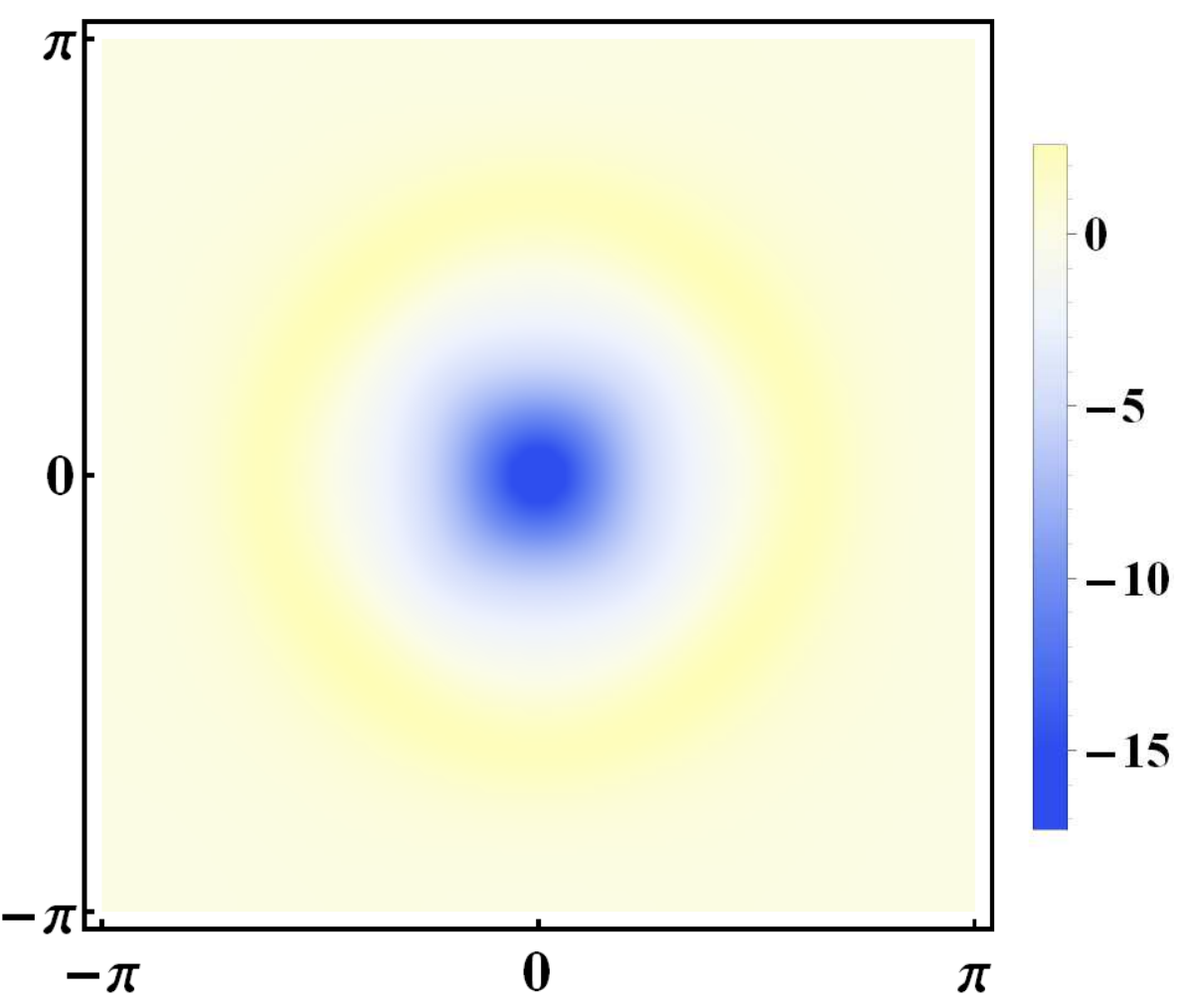} & \includegraphics*[width=0.48\columnwidth]{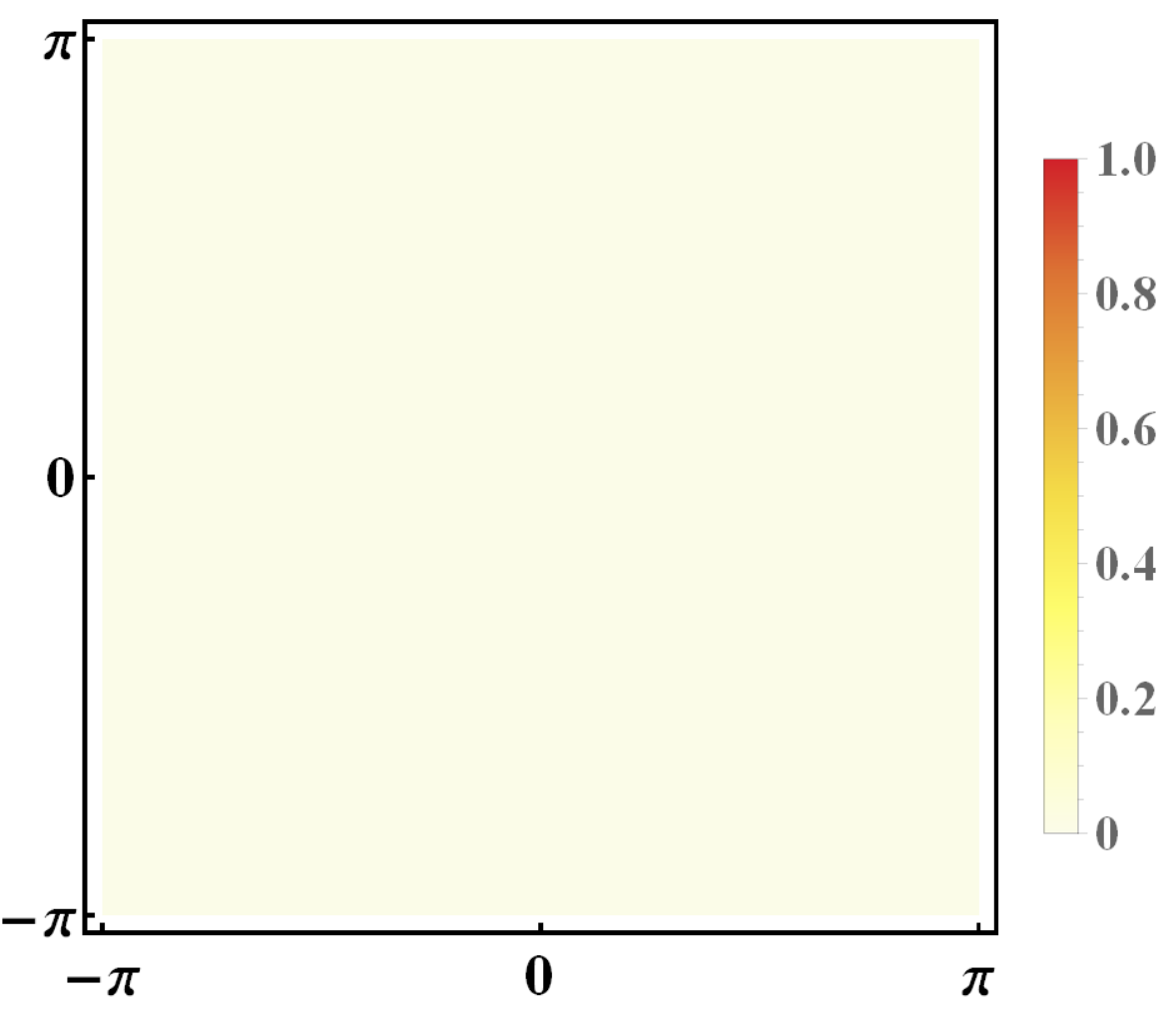} \\
	\end{tabular}
	\caption{(Color online) The real part of the FT of the $S_x$ and $S_y$ components of SP LDOS in arbitrary units for the hole component of a  Shiba state as a function of momentum ($p_x$, $p_y$), for a magnetic impurity with $J_x=2$. 
	We take  
	$\varkappa=0.5$ for the pure p-wave SC with $\dd$ in plane (first row) and $\dd$ perpendicular to the plane (second row).}
	\label{SxMS}
\end{figure}

Our second important result is that the magnitude of the average SP LDOS of the particle and hole component of the SBSs are generically different. Most strikingly, when both the $\dd$ vector and the impurity spin are in-plane (in this case $M^z_\parallel$ is no longer a conserved quantity) 
only two of the four states remain spin-polarized  while the spin polarization of the other 
two goes to zero (see Fig.~\ref{purepJvar}) in the extreme case of a pure p-wave SC.  
This  cancellation can be directly traced back to the orbital nature of the p-wave order parameter which entails that
 $S_x(\mathbf p = \mathbf 0)=S_y(\mathbf p = \mathbf 0)=0$ for this particular SBS components
(see the first row in Fig. \ref{SxMS}).
Despite the fact that this exact cancellation would disappear with the inclusion of the SOC, a strong asymmetry between the SP DOS of the particle and hole components of the SBSs largely survives (see Fig. S7 in the  SI \cite{sm}).

We now focus on the FT of the SP LDOS associated with  SBSs. 
As shown above, the most dramatic situation corresponds to  an in-plane $\dd$ vector.
For a z-impurity, the SP LDOS of the positive energy state (the electron component of the SBS) does not change sign when undergoing the topological transition, while the spin polarization of the negative energy state (the hole component) does (see lower left panel of Fig. \ref{kappavar}). We thus focus on the latter.

\begin{figure}[t]
	\centering
	\begin{tabular}{ccc}
		\multicolumn{3}{c}{\textbf{z-impurity}} \\
		& pure s-wave$\phantom{11}$ & pure p-wave$\phantom{11}$  \\
		\rotatebox{90}{$\phantom{loremipsum} S_z$} &  \includegraphics*[width=0.47\columnwidth]{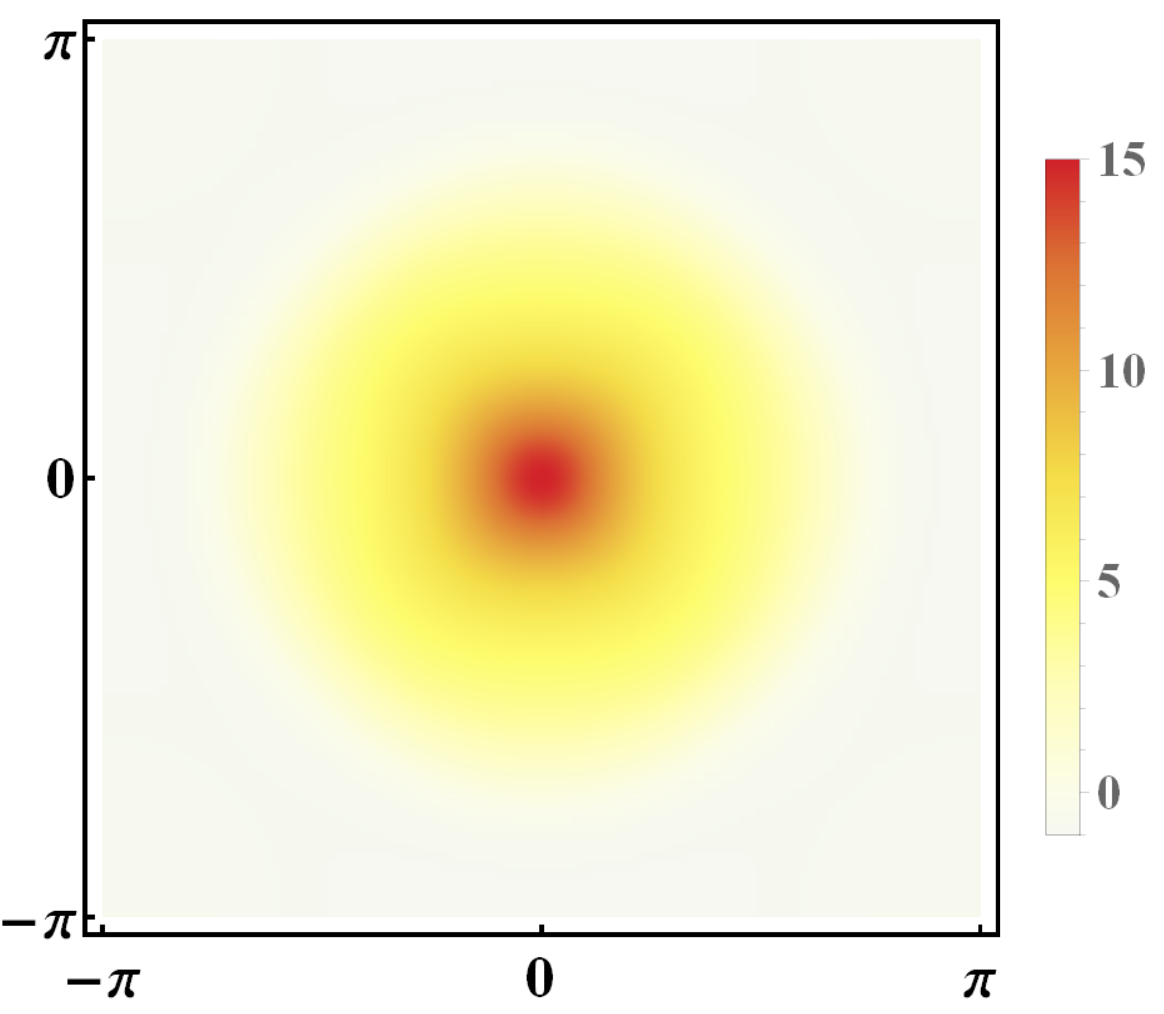} & \includegraphics*[width=0.48\columnwidth]{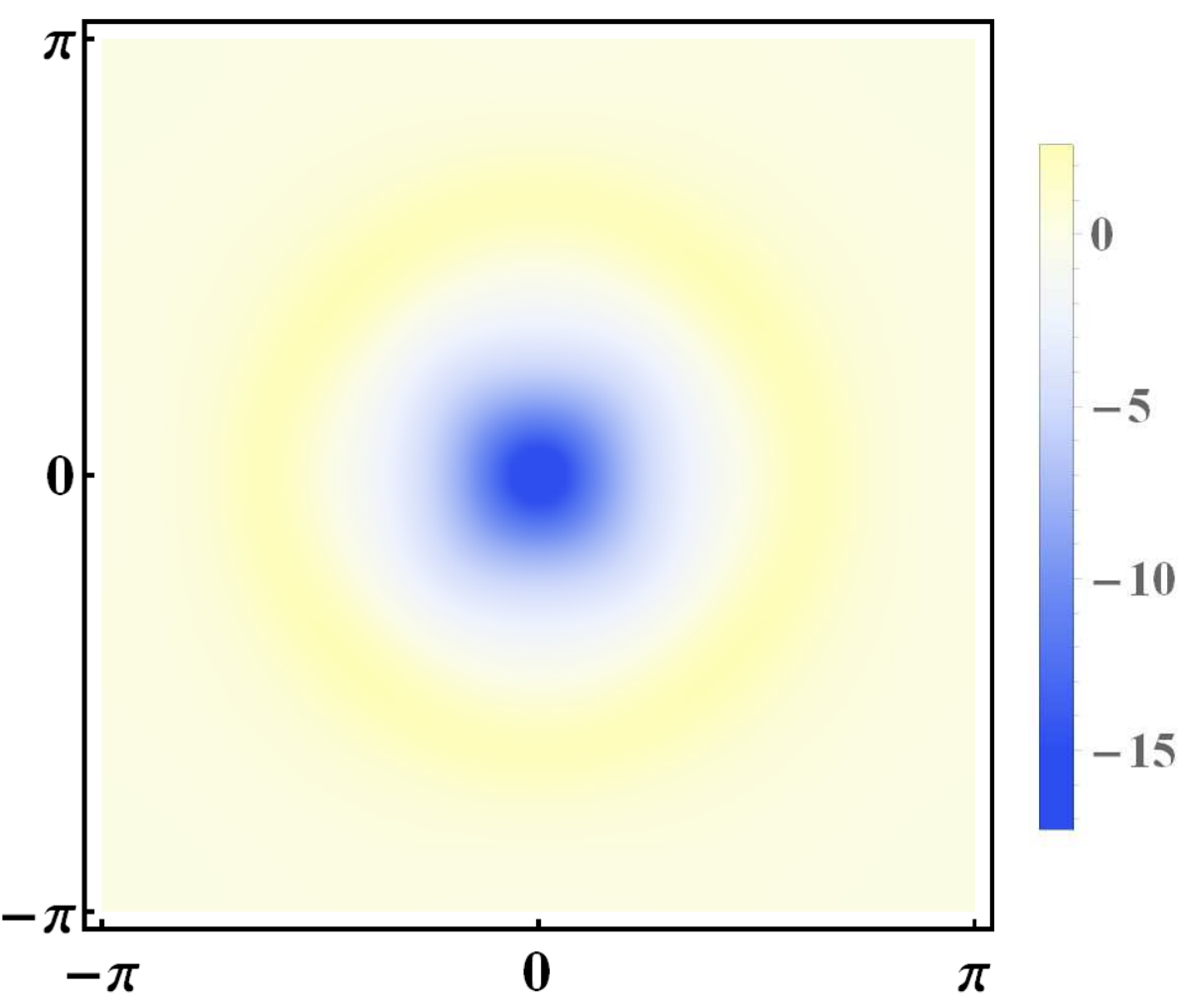}\\
	\end{tabular}
	\caption{(Color online) The real part of the FT of the $S_z$ SP LDOS component for the hole component of a Shiba BS as a function of momentum ($p_x$, $p_y$) for a magnetic impurity with $J_z=2$ and an in-plane $\dd$ vector. We take  $\Delta_s=0.2$, $\varkappa=0$ for a pure s-wave SC, and $\Delta_s=0$, $\varkappa=0.5$ for a pure p-wave SC. }
	\label{SzMS}
\end{figure}

In Fig.~\ref{SzMS} we plot the FT of $S_{z}({\bm p},E)$  (the only non-zero component of the SP LDOS)
for  the pure s-wave (left panel) and pure p-wave (right panel) regimes. Although they have similar shapes, as expected by rotational symmetry along the z-axis,
they show qualitative different behaviors: in the former case, one obtains a central peak and a ring with the same sign, while for the latter case they have opposite signs, corresponding to a spin-flip of the average spin polarization between the dominant s-wave and dominant p-wave regimes. 

Finally, we focus on the FT of the SP LDOS for a state whose total spin polarization goes to zero in the pure p-wave case,  as described in Fig.~\ref{purepJvar}  for both in-plane $\dd$ vector and  spin impurity ($J_x=2$). 
While for the pure s-wave case (not shown) the $S_x$ observable would exhibit the same qualitative features as those described in the left panel of Fig.~\ref{SzMS},
for the pure p-wave SC both $S_x$ and $S_y$ are non-zero  (first row of Fig. \ref{SxMS}), and most strikingly
exhibit characteristic four-fold symmetries. For an in-plane $\dd$-vector, $\mathcal{H}_{imp}$ no longer commutes with $\mathcal{H}_0$
and therefore an in-plane impurity is sensitive to the orbital part of the triplet Cooper pairs.
Due to this four-fold symmetry we have also $S_{x/y}(\mathbf r=\mathbf 0)=0$, in
agreement with analytical solutions (see [\onlinecite{sm}]). 
Note that this is a very unique feature. For comparison we consider also a pure p-wave state with a $\dd_\perp$
vector perpendicular to the plane (second row of Fig. \ref{SxMS}) and we find that in this case only the $S_x$ component of the SP LDOS is non-zero and moreover  has a radial structure as expected by rotational symmetry around the z axis. 
%
Therefore such characteristic spin anisotropy, if detected in  spin-polarized STM, can be used as a signature of the transition into a topological p-dominant regime, as well as an indicator of the direction of the $\dd$ vector. 

We should note that the previous two results are qualitatively unchanged  when comparing s-wave and p-wave dominant SCs instead of pure s-wave and p-wave (see the SI \cite{sm} for a detailed description of the mixed case), though small differences arise such as non-zero values for all the components of the FT of the SP LDOS. Moreover, the results presented above qualitatively hold in the presence of Rashba SOC although the cancellation present in Fig. \ref{purepJvar} becomes only partial 
(see Fig. S7 in the SI \cite{sm}).\\

Note also that the results remain valid if the direction of the impurity spin is arbitrary. As it has already been pointed out in the article, an impurity with magnetic moment along a direction specified by a unit vector $\hat{n}$, gives rise to a total (averaged along the entire space) non-zero polarization only along $\hat{n}$, even if it would give rise to a non-zero spatial spin structure in more than one spin components. Thus, instead of having one non-zero spin-polarised DOS component in a particular direction ($x$, $y$ or $z$), one has  all three of them if all three $J_x, J_y$ and $J_z$ present. We have checked that the average spin structure along a given direction in this case is exactly similar to that corresponding to the impurity spin pointing along this particular direction. For the FT of the SPDOS a particular example is presented in the SI.

\section{Conclusion}
We have analyzed the formation of Shiba states in  2D SCs with broken inversion symmetry with an admixture of s-wave and p-wave superconductivity. We have found that the number, the energy and especially the spin polarization of the Shiba states depend strongly on the ratio between the values of the s-wave and p-wave coupling. We propose to test experimentally the presence of the p-wave coupling, as well as
the direction of the $\dd$ vector with respect to the sample plane by measuring the spin polarization and  energy of the Shiba bound states via spin-polarized STM or transport.  We note that these signatures are only visible in the spin-polarized quantities and in the presence of magnetic impurities. Our results can be tested using quasi-2D SCs such as Sr$_2$RuO$_4$.

{\bf Acknowledgements}
This work is supported by the ERC Starting Independent Researcher Grant NANOGRAPHENE 256965.
We would like to acknowledge interesting discussions with T. Cren, S. Hoffman and T. Ojanen, and financial support from the  French Agence Nationale de la Recherche through the contract Mistral.

\bibliography{biblio_prb}

\newpage
\widetext
\appendix

\section{Determination of the Shiba spectrum  for a pure p-wave superconductor}
We consider a pure p-wave SC with an in-plane d vector $\mathbf{d} = \mathbf{d}_\parallel$.
Using the Nambu basis described in the main text, the low-energy Hamiltonian for a p-wave superconductor is given by:
$$
{\cal H}_0 = \bpm \xi_k & 0 & 0 & i\varkappa k_- \\ 
		 0 & \xi_k & -i\varkappa k_+ & 0 \\
		 0 & i\varkappa k_- & -\xi_k & 0 \\
		 -i\varkappa k_+ & 0 & 0 & -\xi_k 
	\epm,
	\quad k_\pm = k_x \pm i k_y
$$
where $\varkappa$  is the triplet pairing constant, $\xi_k = \frac{k^2}{2m}-\mu$ is the dispersion relation in the continuum limit.
The spectrum of ${\cal H}_0$ is
$
\mathcal{E} = \pm\sqrt{\xi^2_k+\varkappa^2k^2},
$
with a triplet gap parameter
$
\Delta_t = \frac{\varkappa k_F}{\sqrt{1+\varkappa^2}},
$ where we set $v_F=1$. As in the main text, we also set $\hbar=1$.
\\
\\
We consider a single impurity localized at $\mathbf{r}=\mathbf{0}$, with both scalar and magnetic components of the potential  $(U, {\bm J})$, and ${\bm J} = J \left( \sin\theta \cos\phi, \sin\theta \sin\phi, \cos\theta \right)$, where $U$ and $J$ are the scalar and magnetic strengths respectively. The corresponding Hamiltonian can be written as:
$$
\mathcal{H}_{imp} = \delta(\mathbf{r})
					\bpm 
						U+J\cos\theta & J \sin\theta e^{-i\phi} & 0 & 0 \\
						J \sin\theta e^{i\phi} & U-J \cos\theta & 0 & 0 \\
						0 & 0 & -U+J\cos\theta & J \sin\theta e^{-i\phi} \\
						0 & 0 & J \sin\theta e^{i\phi} & -U-J\cos\theta
					\epm \equiv \delta(\mathbf{r}) \cdot V.				
$$
\\
To find the energy levels of the Shiba states induced by the impurity we follow a standard procedure detailed in \cite{Pientka2013}. The unperturbed Green's function in momentum space is:
$$
G_0(E,\mathbf{k}) \equiv \left(E-{\cal H}_0 \right)^{-1} = -\frac{1}{\xi_k^2+\varkappa^2 k^2 - E^2}
		\bpm 	E+\xi_k & 0 & 0 & i\varkappa k_- \\ 
				0 & E+\xi_k & -i\varkappa k_+ & 0 \\
		 		0 & i\varkappa k_- & E-\xi_k & 0 \\
		 		-i\varkappa k_+ & 0 & 0 & E-\xi_k 
		\epm,
$$
To obtain $G_0(E,\mathbf{r}=\mathbf{0})$ we integrate $G_0(E,\mathbf{k})$ over all momenta, thus we need to perform the following integrals: 
\begin{eqnarray*}
X_0 &=& -\int \frac{d\mathbf{k}}{(2\pi)^2} \frac{E}{\xi_k^2+\varkappa^2 k^2 - E^2} \\
X_1 &=& -\;\mathrm{v.p.}\negthickspace\int \frac{d\mathbf{k}}{(2\pi)^2} \frac{\xi_k}{\xi_k^2+\varkappa^2 k^2 - E^2} \\
X^\pm_2 &=& \mp \negthickspace\int \frac{d\mathbf{k}}{(2\pi)^2} \frac{i \varkappa k_\pm}{\xi_k^2+\varkappa^2 k^2 - E^2} \\
\end{eqnarray*}
The last integral is zero due to the angular part. The second integral has a UV divergence thus we need to use a natural cut-off which, in this particular case, is equivalent to computing the principal value of the integral. We linearise $\xi_k$ around the Fermi level, and using the spherical symmetry of the integrals we change variables:
$$
\xi_k \approx v_F (k-k_F), \quad \int \frac{d\mathbf{k}}{(2\pi)^2} = \nu \int d\xi_k, \;\text{where}\; \nu = \frac{m}{2\pi}
$$
Performing the integrals we obtain:
$$
X_0 = -\frac{\pi \nu}{\sqrt{1+\varkappa^2}} \frac{E}{\sqrt{\Delta_t^2-E^2}}, \quad X_1 = \frac{\pi \nu}{\sqrt{1+\varkappa^2}} \frac{\Delta_t}{\sqrt{\Delta_t^2-E^2}} \frac{\varkappa}{\sqrt{1+\varkappa^2}}, \quad X^\pm_2 = 0
$$
The Green's function for $\mathbf{r}=\mathbf{0}$ then takes the form:
\begin{equation*}
G_0(\epsilon,\mathbf{r}=\mathbf{0}) = -\frac{\pi\nu}{\sqrt{1+\varkappa^2}} \frac{1}{\sqrt{1-\epsilon^2}}
		\bpm 	\epsilon-\gamma & 0 & 0 & 0 \\ 
				0 & \epsilon-\gamma & 0 & 0 \\
		 		0 & 0 & \epsilon+\gamma & 0 \\
		 		0 & 0 & 0 & \epsilon+\gamma 
		\epm, \quad\text{where}\; \epsilon=\frac{E}{\Delta_t},\;\gamma = \frac{\varkappa}{\sqrt{1+\varkappa^2}}.
\end{equation*}
The eigenenergies and corresponding eigenvectors can be found using $\left[ \mathbb{I}_4-V G_0(\epsilon,\mathbf{r}=\mathbf{0}) \right] \psi(0) = 0$.  The electron component of the spin operators corresponds in our Nambu basis to 
$\hat{S}_i=\frac{1-\tau_z}{2}\otimes\sigma_i$ with $i=x,y,z$.\\

To illustrate the method, we  consider an impurity with a mixed scalar  and magnetic impurity potential along the z-axis.
$$
V = 	\bpm 	
						U+J_z & 0 & 0 & 0 \\
						0 & U-J_z & 0 & 0 \\
						0 & 0 & -U+J_z & 0 \\
						0 & 0 & 0 & -U-J_z
					\epm.
$$
The eigenvalues and eigenvectors can be found from the equation:
$$
\left[ \mathbb{I}_4+\frac{\pi \nu}{\sqrt{1-\epsilon^2}}
		\bpm 	(U+J_z)(\epsilon-\gamma) & 0 & 0 & 0 \\ 
				0 & (U-J_z)(\epsilon-\gamma) & 0 & 0 \\
		 		0 & 0 & -(U-J_z)(\epsilon+\gamma) & 0 \\
		 		0 & 0 & 0 & -(U+J_z)(\epsilon+\gamma) 
		\epm \right] \psi(0) = 0.
$$
Introducing two dimensionless parameters $\beta_\pm = \frac{\pi \nu (U \pm J_z)}{\sqrt{1+\varkappa^2}}$, we get for the case  $\beta_->0$:
\begin{eqnarray*}
\epsilon_{\bar 1} &=& -\frac{-\gamma \beta_-^2 + \sqrt{1+\beta_-^2 (1-\gamma^2)}}{1+\beta_-^2}, \quad \psi_{\bar 1}(0)=\bpm 0 \\ 1 \\ 0 \\0 \epm,\quad \langle S_x\rangle_{\bar 1} = \langle S_y\rangle_{\bar 1} = \langle S_z\rangle_{\bar 1} = 0 \\
\epsilon_{\bar 2} &=& - \frac{-\gamma \beta_+^2 + \sqrt{1+\beta_+^2 (1-\gamma^2)}}{1+\beta_+^2}, \quad \psi_{\bar 2}(0)=\bpm 1 \\ 0 \\ 0 \\0 \epm,\quad \langle S_x\rangle_{\bar 2} = \langle S_y\rangle_{\bar 2} = \langle S_z\rangle_{\bar 2} = 0  \\
\epsilon_2 &=& +\frac{-\gamma \beta_+^2 + \sqrt{1+\beta_+^2 (1-\gamma^2)}}{1+\beta_+^2}, \quad \psi_2(0)=\bpm 0 \\ 0 \\ 0 \\1 \epm,\quad \langle S_x\rangle_{2} = \langle S_y\rangle_{2} = 0, \langle S_z\rangle_{2} = -1\\
\epsilon_1 &=& +\frac{-\gamma \beta_-^2 + \sqrt{1+\beta_-^2 (1-\gamma^2)}}{1+\beta_-^2}, \quad \psi_1(0)=\bpm 0 \\ 0 \\ 1 \\0 \epm,\quad \langle S_x\rangle_{1} = \langle S_y\rangle_{1} = 0, \langle S_z\rangle_{1} = 1,
\end{eqnarray*}
whereas for the case of $\beta_-<0$ we obtain:
\begin{eqnarray*}
\epsilon_{\bar 1} &=& -\frac{\gamma \beta_-^2 + \sqrt{1+\beta_-^2 (1-\gamma^2)}}{1+\beta_-^2}, \quad \psi_{\bar 1}(0)=\bpm 0 \\ 0 \\ 1 \\0 \epm,\quad \langle S_x\rangle_{\bar 1} = \langle S_y\rangle_{\bar 1} = 0,\; \langle S_z\rangle_{\bar 1} = 1\\
\epsilon_{\bar 2} &=& - \frac{-\gamma \beta_+^2 + \sqrt{1+\beta_+^2 (1-\gamma^2)}}{1+\beta_+^2}, \quad \psi_{\bar 2}(0)=\bpm 1 \\ 0 \\ 0 \\0 \epm,\quad \langle S_x\rangle_{\bar 2} = \langle S_y\rangle_{\bar 2} = \langle S_z\rangle_{\bar 2} = 0 \\
\epsilon_2 &=& + \frac{-\gamma \beta_+^2 + \sqrt{1+\beta_+^2 (1-\gamma^2)}}{1+\beta_+^2}, \quad \psi_2(0)=\bpm 0 \\ 0 \\ 0 \\1 \epm,\quad \langle S_x\rangle_2 = \langle S_y\rangle_2 = 0, \; \langle S_z\rangle_2 = -1 \\
\epsilon_1 &=& +\frac{\gamma \beta_-^2 + \sqrt{1+\beta_-^2 (1-\gamma^2)}}{1+\beta_-^2}, \quad \psi_1(0)=\bpm 0 \\ 1 \\ 0 \\0 \epm,\quad \langle S_x\rangle_1 = \langle S_y\rangle_1 = \langle S_z\rangle_1 = 0 
\end{eqnarray*}

\section{Degeneracy lifting for a scalar impurity in a pure p-wave SC}
\begin{figure}[h]
\noindent\includegraphics[width=7cm,height=4cm]{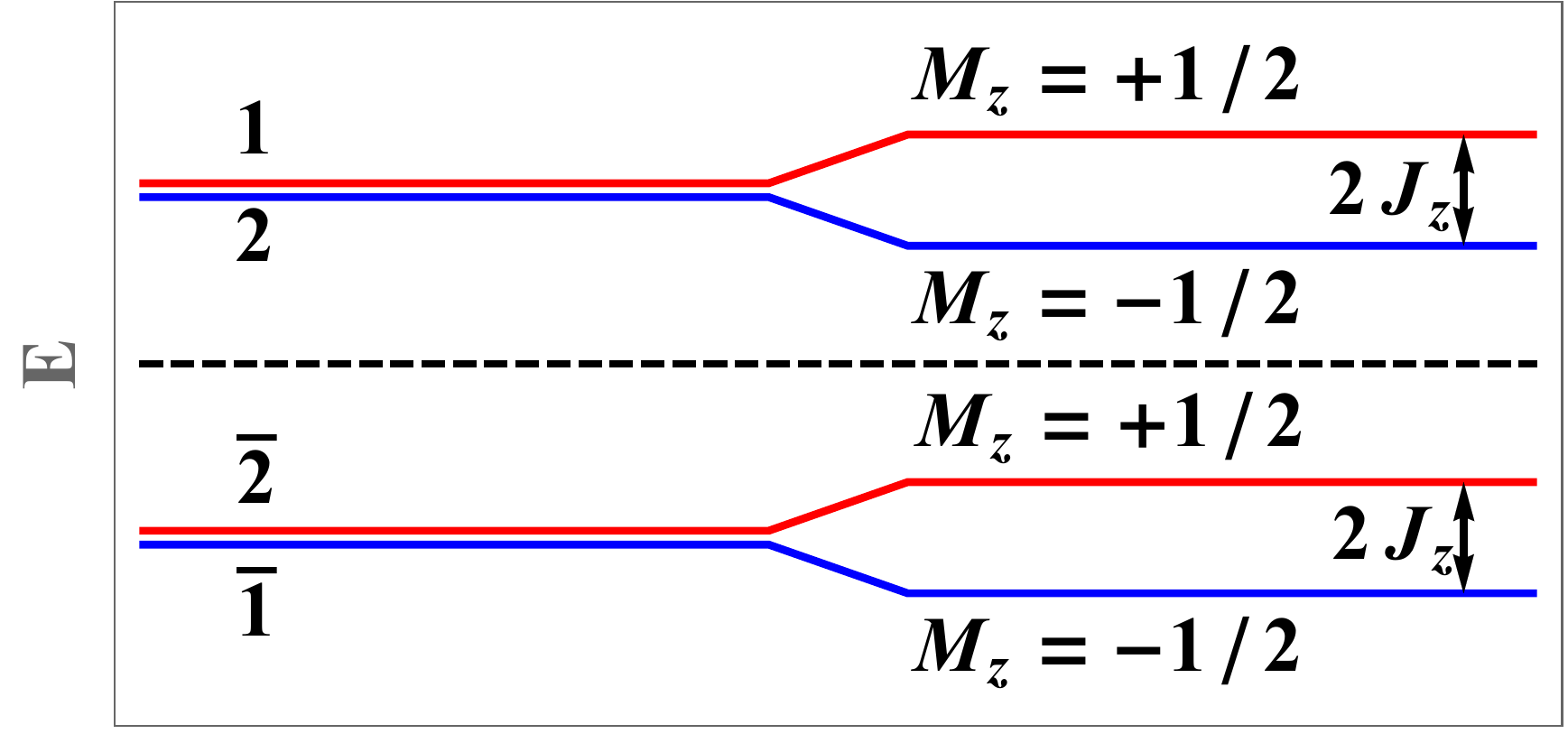}\hskip 1cm
\noindent\includegraphics[width=7cm,height=4cm]{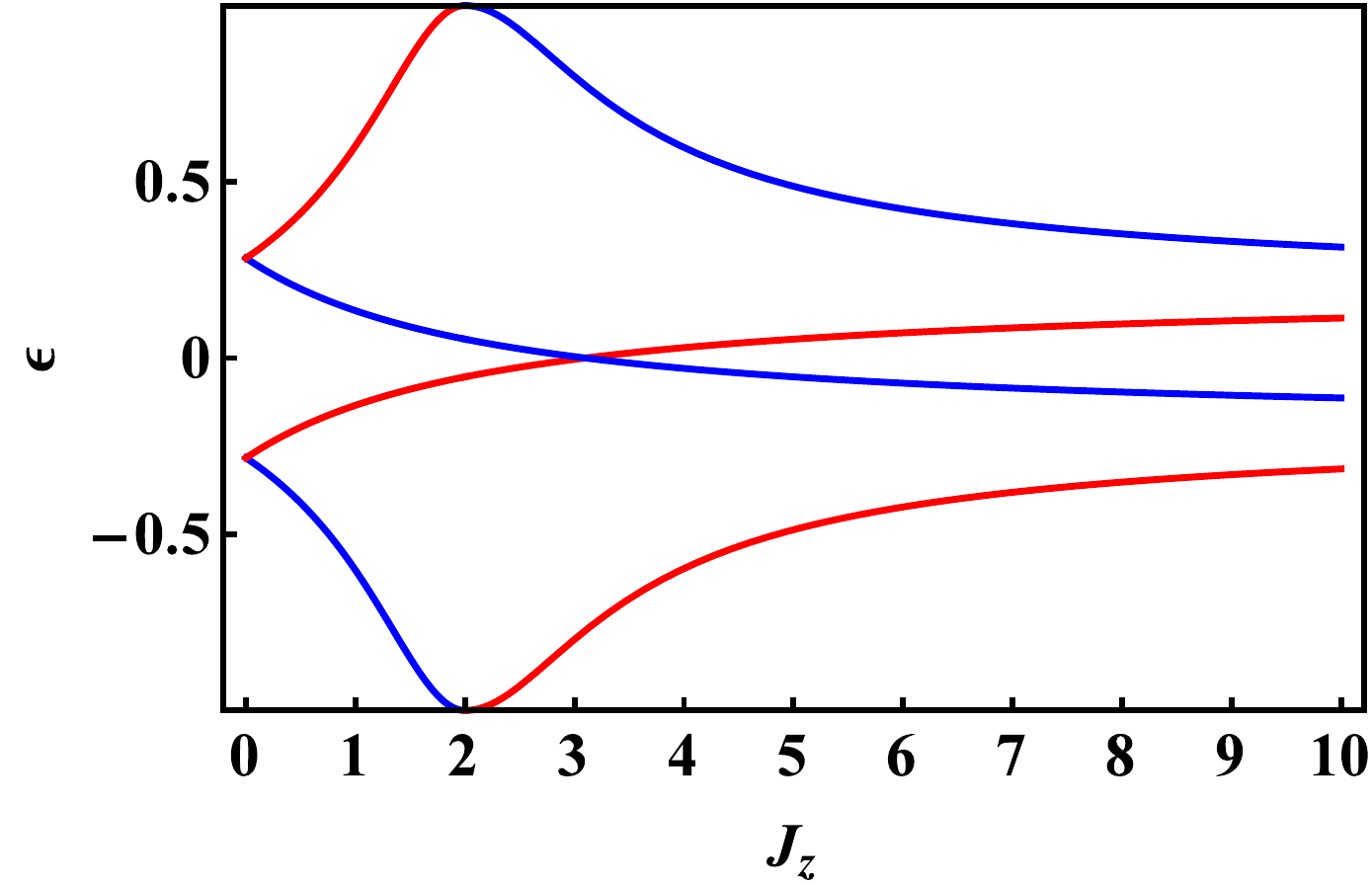}
	\caption{(Color online) Left: Sketch of the energy levels obtained by perturbation theory in  impurity strength $J_z$. The colours stand for the angular momentum: red for $+1/2$ and blue for $-1/2$.
	Right:
	Energy levels as a function of impurity strength $J_z$  obtained analytically.  We set $\varkappa = 0.2$ and used the same colour code.}
	\label{Shibamix}
\end{figure}

A scalar impurity in a pure p-wave SC has two degenerate Shiba states. For the in-plane $\bm d$ vector case which is TRS, the two SBS form Kramers pairs with total angular momentum $M_z=M^z_\parallel=\pm 1/2$ (for an out-of-plane $\bm d$ vector, the  Shiba states can be labeled with $s_z=\pm 1/2$). When a small growing magnetic moment is introduced in the impurity potential, this degeneracy
is lifted and the two levels split. For a magnetic impurity along the z-axis, $M_z=\pm 1/2$ is conserved and the SBS with $M_z=-1/2$ has a lower energy due to antiferromagnetic interactions (see Fig. \ref{Shibamix} for a sketch of the energy levels of Shiba bound states).
 Note that an in-gap quasi-particle excitation with $M_z=-1/2$ is {\it a priori} a coherent superposition of states with $(L^z=0,s^z=-1/2)$ and $(L^z=-1,s^z=1/2)$, the former being more electron-like, the latter being more hole-like. Since the impurity potential  is point-like, only the $(L_z=0,s^z=\pm 1/2)$ components are affected by the impurity which explains the labelling used in Fig. \ref{Shibamix}.
 It is easy to compute this splitting using a series representation for $\gamma=J_z/U \ll 1$. We find
$$
\delta \epsilon = \epsilon_{\bar 2}-\epsilon_{\bar 1} \approx \frac{2 \beta ^2}{(1+\beta^2)^2} \left[1+\frac{\gamma}{\sqrt{1+\beta^2(1-\gamma^2)}} \right] \gamma,
$$
where $\beta = \frac{\pi \nu U}{\sqrt{1+\varkappa^2}}$.
We plot the energy levels using both the analytical expressions given above, and numerical calculations performed using the T-matrix approximation and a discretized (tight-binding) Hamiltonian. The evolution of the energy levels as a function of the dimensionless magnetic impurity strength is ploted  in Fig. \ref{Shibamix}.

\section{Derivation of the effective gaps for a superconductor with mixed singlet and triplet pairing}

Below we consider a SC with mixed singlet and triplet pairing, defined by an in-plane $\dd$ vector $\mathbf{d} = \mathbf{d}_\parallel$. The corresponding Hamiltonian and the spectrum are given by Eqs.~(\ref{h0}) and (\ref{spspectrum}) respectively. Note that in all the further calculations we  linearize the dispersion relation around the Fermi energy. All the integrations are performed using this approximation.

The unperturbed Green's function in momentum space can be written as:
$$
G_0(E,{\bm k}) = \frac{1}{2} \sum\limits_{\sigma=\pm} G^\sigma_0(E,{\bm k}),
$$
where
$$
G^\sigma_0(E,{\bm k}) = -\frac{1}{\xi_{{\bm k}}^2+(\Delta_s + \sigma\varkappa k)^2 - E^2}
		\bpm 	1 & i \sigma e^{-i\varphi} \\ 
				-i \sigma e^{i\varphi} & 1
		\epm \otimes	
		\bpm 	E+\xi_{{\bm k}} & \Delta_s + \sigma\varkappa k \\ 
				\Delta_s + \sigma\varkappa k & E-\xi_{{\bm k}}
		\epm,
$$
To get the zero coordinate value we linearize $\xi_{\bm k}$ around Fermi level (see Appendix A) and calculate the following integrals:
$$
G^\sigma_0(E,\mathbf{r}=\mathbf{0}) = -\int\frac{d{\bm k}}{(2\pi)^2}\frac{1}{\xi_{{\bm k}}^2+(\Delta_s + \sigma\varkappa k)^2 - E^2}
		\bpm 	1 & 0 \\ 
				0 & 1
		\epm \otimes	
		\bpm 	E+\xi_{{\bm k}} & \Delta_s + \sigma\varkappa k \\ 
				\Delta_s + \sigma\varkappa k & E-\xi_{{\bm k}}
		\epm,
$$
We need to calculate integrals of the form 
$$
I_{n}=-\int\frac{d{\bm k}}{(2\pi)^2}\frac{ A_n}{\xi_k^2+(\Delta_s + \sigma\varkappa k)^2 - E^2},
$$
where $n=1,2,3$ and $A_1=E$, $A_2=\xi_k$, and $A_3=\Delta_s+\sigma\varkappa k $.
Using $k=k_F + \xi_{{\bm k}}/v_F$ we get:
$$
-\int\frac{d{\bm k}}{(2\pi)^2}\frac{A_n}{\xi_k^2+(\Delta_s + \sigma\varkappa k)^2 - E^2} = -\frac{\nu}{1+\varkappa^2} \int d\xi_k \frac{A_n}{\left(\xi_k + \sigma \gamma \Delta^\sigma_{eff} \right)^2 + \omega_\sigma^2},
$$
where
$
\nu \equiv \frac{m}{2\pi},\; \gamma \equiv \frac{\varkappa}{\sqrt{1+\varkappa^2}},\;\omega^2_\sigma \equiv \frac{\left(\Delta^\sigma_{eff}\right)^2 - E^2}{1+\varkappa^2},\; \Delta^\sigma_{eff} \equiv \frac{\Delta_s + \sigma \varkappa k_F}{\sqrt{1+\varkappa^2}}.
$
Performing the integrations we get:
\begin{align*}
I_1 &=  -\frac{\pi \nu}{\sqrt{1+\varkappa^2}} \frac{E}{\sqrt{\left(\Delta^\sigma_{eff}\right)^2-E^2}} \\
I_2 &=  +\frac{\pi \nu}{\sqrt{1+\varkappa^2}} \frac{\Delta^\sigma_{eff}}{\sqrt{\left(\Delta^\sigma_{eff}\right)^2-E^2}} \sigma \gamma\\
I_3 &=  -\frac{\pi \nu}{\sqrt{1+\varkappa^2}} \frac{\Delta^\sigma_{eff}}{\sqrt{\left(\Delta^\sigma_{eff}\right)^2-E^2}} \frac{1}{\sqrt{1+\varkappa^2}}
\end{align*}
The integrals of all off-diagonal components are zeros due to the angular part. Thus we get:
$$
G^\sigma_0(E,\mathbf{r}=\mathbf{0}) = -\frac{\pi \nu}{\sqrt{1+\varkappa^2}} \frac{1}{\sqrt{\left(\Delta^\sigma_{eff}\right)^2-E^2}}
\bpm
	E - \sigma \gamma \Delta^\sigma_{eff} & 0 & \frac{\Delta^\sigma_{eff}}{\sqrt{1+\varkappa^2}} & 0 \\
	0 & E - \sigma \gamma \Delta^\sigma_{eff} & 0 & \frac{\Delta^\sigma_{eff}}{\sqrt{1+\varkappa^2}} \\
	\frac{\Delta^\sigma_{eff}}{\sqrt{1+\varkappa^2}} & 0 & E + \sigma \gamma \Delta^\sigma_{eff} & 0 \\
	0 & \frac{\Delta^\sigma_{eff}}{\sqrt{1+\varkappa^2}} & 0 & E + \sigma \gamma \Delta^\sigma_{eff}
\epm.
$$
It is clear that all the features arising from the bulk of a SC with mixed s-wave and p-wave types of pairing are originating from the expressions for the effective gaps, namely:
$$
\Delta^\pm_{eff}= \frac{|\Delta_s \pm \varkappa k_F|}{\sqrt{1+\varkappa^2}},
$$
the smallest of the two being the superconducting gap.
\end{document}